\documentclass{aa} % for a referee version
\usepackage{graphicx, pstricks}
\usepackage{amsmath}
\usepackage{txfonts}
\usepackage{tikz}
\usepackage{xspace}
\usepackage{subfigure}
\usepackage[draft]{hyperref}
\usepackage{algorithm} 
\usepackage{algpseudocode}

\newcommand{\nuna}[2]{(#1)\,#2} % For asteroid number-name pairs
\newcommand{\astorb}{\texttt{astorb}\xspace}
\newcommand{\cometpro}{\texttt{cometpro}\xspace}
\newcommand{\herschel}{\textsl{Herschel}\xspace}

\tikzset{
  treenode/.style = {shape=rectangle, rounded corners,
                     draw, align=center,
                     top color=white,
                     bottom color=blue!20,
                     font=\ttfamily\normalsize},
  root/.style     = {treenode,% font=\Large,
                     bottom color=blue!30},
  env/.style      = {treenode, font=\ttfamily\normalsize},
  dummy/.style    = {circle,draw}
}

\usepackage{hyperref}
\hypersetup{
    bookmarks=true,         % show bookmarks bar?
    unicode=true,           % non-Latin characters in Acrobat's bookmarks
    pdftoolbar=true,        % show Acrobat's toolbar?
    pdfmenubar=true,        % show Acrobat's menu?
    pdffitwindow=true,      % page fit to window when opened
    pdftitle={ESASky \& SSOs},
    pdfauthor={Racero et al.},
    pdfsubject={Planetary Science},
    pdfkeywords={},         % list of keywords
    pdfnewwindow=true,      % links in new window
    colorlinks=true,        % false: boxed links; true: colored links
    linkcolor=gray,         % color of internal links
    citecolor=blue,         % color of links to bibliography
    filecolor=gray,         % color of file links
    urlcolor=gray           % color of external links
} 

\date{Received March 26, 2021; accepted August 16, 2021}
\begin{document} 

   \title{ESASky SSOSS: Solar System Object Search Service and the case of Psyche}

   \subtitle{}

   \author{E. Racero\inst{1,2}, 
                F. Giordano\inst{2,3}, 
                B. Carry\inst{4,5},
                J. Berthier\inst{5},
                T. M{\"u}ller\inst{6},
                M. Mahlke\inst{4},
                I. Valtchanov\inst{7},
                D. Baines\inst{7}, 
            S. Kruk\inst{8},
                B. Mer\'{\i}n\inst{9},
                S. Besse\inst{9}, 
                M. K{\"u}ppers\inst{9},
                E. Puga\inst{9},
            J. Gonz\'alez N\'u\~nez\inst{2}, 
                P. Rodr\'iguez\inst{2},
                I. de la Calle \inst{7},
        B. L\'opez-Marti\inst{10},
        H. Norman\inst{11},
        M. W\r{a}ngblad\inst{11},
        M. L\'opez-Caniego\inst{12}
        \and
         N. \'Alvarez Crespo \inst{9}
}

   \institute{Departamento de Astrof\'isica y CC. de la Atm\'osfera, Facultad de CC. F\'isicas, Avenida Complutense s/n, E-28040 Madrid, Spain
    \and
    Serco for the European Space Agency (ESA), European Space Astronomy Centre (ESAC), Camino Bajo del Castillo s/n, 28692 Villanueva de la Ca\~nada, Madrid, Spain\\
    \email{eracero@sciops.esa.int}
    \and
    Departamento de Fisica Teorica, Universidad Autonoma de Madrid, Cantoblanco, 28049 Madrid, Spain
        \and
        Universit\'e C\^ote d'Azur, Observatoire de la C\^ote d'Azur, CNRS, Laboratoire Lagrange, France
        \and
        Institut de M\'ecanique C\'eleste et de Calcul des \'Eph\'em\'erides, Observatoire de Paris, UMR8028 CNRS, 77 av. Denfert-Rochereau, 75014 Paris, France
        \and
        Max-Planck-Institut f\"ur extraterrestrische Physik (MPE), Giessenbachstrasse 1, 85748 Garching, Germany
        \and
    Telespazio UK for the European Space Agency (ESA), European Space Astronomy Centre (ESAC), Camino Bajo del Castillo s/n, 28692 Villanueva de la Ca\~nada, Madrid, Spain
        \and
        European Space Agency (ESA), European Space Research and Technology Centre (ESTEC), Keplerlaan 1, 2201 AZ Noordwijk, The Netherlands
    \and
    European Space Agency (ESA), European Space Astronomy Centre (ESAC), Camino Bajo del Castillo s/n, 28692 Villanueva de la Ca\~nada, Madrid, Spain
        \and
        Centro de Astrobiología (CSIC-INTA), Campus ESAC (ESA),  E-28692 Villanueva de la Ca{\~n}ada (Madrid), Spain
    \and
    Winter Way AB, S\"{o}dra L\r{a}ngv{\"a}gen 27A, 443 38, Lerum, Sweden
    \and
    Aurora for ESA, European Space Astronomy Centre, Operations Department, 28691 Villanueva de la Ca\~nada, Spain
    }

\abstract
{The store of data collected in public astronomical archives across the world is continuously expanding and, thus, providing a convenient interface for accessing this information is a major concern  for ensuring a second life for the data. In this context, Solar System Objects (SSOs) are often difficult or even impossible to query, owing to their ever-changing sky coordinates.}
{Our study is aimed at providing the scientific community with a search service for all potential detections of SSOs among the ESA astronomy archival imaging data, called the Solar System Object Search Service (SSOSS). We illustrate its functionalities using the case of asteroid (16) Psyche, for which no information in the far-IR (70-500 $\mu$m) has previously been reported, to derive its thermal properties in preparation for the upcoming NASA Psyche mission.}
{We performed a geometrical cross-match of the orbital path of each object, as seen by the satellite reference frame, with respect to the public high-level imaging footprints stored in the ESA archives. There are about 800,000 asteroids and 2,000 comets included in the SSOSS, available through ESASky, providing both targeted and serendipitous observations. For this first release, three missions were chosen: XMM-Newton, the Hubble Space Telescope (HST), and \herschel.}
{We present a catalog listing all potential detections of asteroids within estimated limiting magnitude or flux limit in Herschel, XMM-Newton, and HST archival imaging data, including 909 serendipitous detections in Herschel images, 985 in XMM-Newton Optical Monitor camera images, and over 32,000 potential serendipitous detections in HST images. We also present a case study: the analysis of the thermal properties of Psyche from four serendipitous Herschel detections, combined with previously published thermal IR measurements. We see strong evidence for an unusual drop in (hemispherical spectral) emissivity, from 0.9 at 100 $\mu$m down to about 0.6 at 350 $\mu$m, followed by a possible but not well-constrained increase towards 500 $\mu$m, comparable to what was found for Vesta. The combined thermal data set puts a strong constraint on Psyche’s thermal inertia (between 20 to 80\,J\,m$^{-2}$\,s$^{-1/2}$\,K$^{-1}$) and favors an intermediate to low level surface roughness (below 0.4 for the rms of surface slopes).}
{Using the example of Psyche, we show how the SSOSS provides fast access to observations of SSOs from the ESA astronomical archives, regardless of whether the particular object was the actual target. This greatly simplifies the task of searching, identifying, and retrieving such data for scientific analysis.}
\titlerunning{ESASky SSOS: The Case of Psyche}
\authorrunning{Racero et al.}
  \maketitle

%----------------------------------------------------------------------------------------------------%
%----------------------------------------------------------------------------------------------------%
\section{Introduction}

Providing the Solar System research community with swift and easy access to the astronomical data archives is a long-standing issue. Moreover, the consistently increasing store of archival data coming from a variety of facilities, both from ground-based telescopes and space missions, has led to the need for single points of entry for exploration purposes. However, moving targets seen at different sky positions and under very different observing geometries are not easy to aggregate within a single tool.

In general, multi-epoch observations over several oppositions are required to compute the orbit to a sufficient level of accuracy required for targeted studies. The archived imaging data contains both serendipitous and targeted observations of asteroids, where, in particular, the astrometry of the former can greatly reduce these ephemerides uncertainties at visible wavelengths when harvested \citep[e.g., precovery of near-Earth asteroids][]{2014-AN-335-Solano}. Moreover, the extracted photometry can be used to constrain the phase function and, if multi-band observations were acquired within a short period of time, they can even allow for a color determination and rudimentary taxonomic classification of the asteroid \citep{2013-Icarus-226-DeMeo, 2016-PSS-shevchenko}. 

Within this context, the ESAC Science Data Centre (ESDC)\footnote{\url{http://cosmos.esa.int/web/esdc}}, located at the European Space Astronomy Centre (ESAC) has developed ESASky \citep{2018A&C....24...97G}, a science-driven discovery portal for exploring the multi-wavelength sky, providing a fast and intuitive access to all ESA astronomy archive holdings. Released in May 2016, ESASky\footnote{\url{https://sky.esa.int}} is a Web application that sits on top of ESAC hosted archives, with the goal of serving as an interface to all high-level science products generated by ESA astronomy missions. The data spans from radio to gamma-ray regimes, and includes the Planck, \herschel, ISO, HST, XMM-Newton and INTEGRAL missions. In addition, ESASky provides access to data from other international space agency missions (e.g., Chandra from NASA and AKARI and Suzaku from JAXA) and provides access to data from major astronomical data centers and observatories, such as the European Southern Observatory (ESO), the Canadian Astronomy Data Center (CADC), and the Mikulski Archive for Space Telescopes (MAST). ESASky is designed to be exceptionally visual \citep{2017-PASP-129-Baines}, allowing users to: see where in the sky all missions and observatories have observed; find all available data for their targets of interest; overlay catalogue data; visualize which objects have associated publications; perform initial planning of James Webb Space Telescope observations; and change the background all-sky images (HiPS; \citealt{2015A&A...578...A114}) from many different missions and observatories. However, a clear interface with the Solar System community in terms of the scientific exploitation of these astronomical data holdings is not typically accessible.

Efforts to tackle this issue are already in place, such as the Solar System Object Image Search by the Canadian Astronomy Data Centre (CADC) \footnote{\url{http://www.cadc-ccda.hia-iha.nrc-cnrc.gc.ca/en/ssois/}} \citep{2012-PASP-124-Gwyn} and SkyBoT\footnote{\url{http://vo.imcce.fr/webservices/skybot/}}\citep[IMCCE,][]{2006ASPC..351..367B, 2008-ACM-Berthier, 2009-EPSC-Berthier, 2016-MNRAS-458-Berthier}, plus a number of ephemeris services, such as
Horizons\footnote{\url{https://ssd.jpl.nasa.gov/horizons.cgi}} (NASA/JPL), Miriade\footnote{\url{http://vo.imcce.fr/webservices/miriade/}}, and the Minor Planet \& Comet Ephemeris Service\footnote{\url{https://www.minorplanetcenter.net/iau/mpc.html}} (MPC).

In this first integration of the Solar System Object Search Service (SSOSS), we enable users to discover all targeted and serendipitous observations of a given SSO present in the ESA \herschel, HST, and XMM-Newton archives. Upon user input, the official designation of the target is first resolved using the SsODNet\footnote{\url{http://vo.imcce.fr/webservices/ssodnet/}} service. Then the search engine retrieves all the pre-computed results for all the observations matching the input SSO provided. These results are pre-computed as a geometrical cross-match between the observation footprints and the ephemerides of the SSOs within the exposure time frame of the observations and stored in the service-dedicated database schema.

To showcase the capabilities of this tool, the flux values at 70$\,\mu$m and 160$\,\mu$m of asteroid \nuna{16}{Psyche} are reported from four serendipitous detections in the \herschel Space Observatory archival data. These are the first detections of this object in this energy regime, making these results of significant importance for the upcoming NASA Discovery \textsl{Psyche} mission, expected to be launched in 2022 to visit the asteroid in 2026.

The article is organized as follows: Section~\ref{inputdata} describes the input sample of small bodies and the archival high-level metadata products used in this work. Section~\ref{sec:methods} presents the pipeline software implementation and algorithms involved. In Section~\ref{sec:results}, the output catalogues for each potential detection and discussion of results are presented. The far infrared photometry and thermal modeling of asteroid Psyche is included in Section~\ref{sec:psyche}. Finally, our conclusions and plans for future works are described in Section~\ref{sec:conclusions}.

%----------------------------------------------------------------------------------------------------%
\section{Inputs}\label{inputdata}

\subsection{Samples of Solar System Objects}

The input asteroid catalogue was retrieved from the Lowell Observatory Asteroid Orbital Parameter database (\astorb\footnote{\url{http://asteroid.lowell.edu}}). This database is continuously changing and growing, so this work is based on the snapshot taken on July 5, 2019, containing 795,673 objects.

This catalogue was selected based on the availability of the current position uncertainty (CEU) parameter, allowing for the propagation of uncertainties in time and thus permitting us to provide positional uncertainties for each potential detection, while also taking them into account for the cross-match computation, as explained in Section~\ref{geom_xmatch}.

\begin{figure}[!ht]
  \centering
  \includegraphics[width=.45\textwidth]{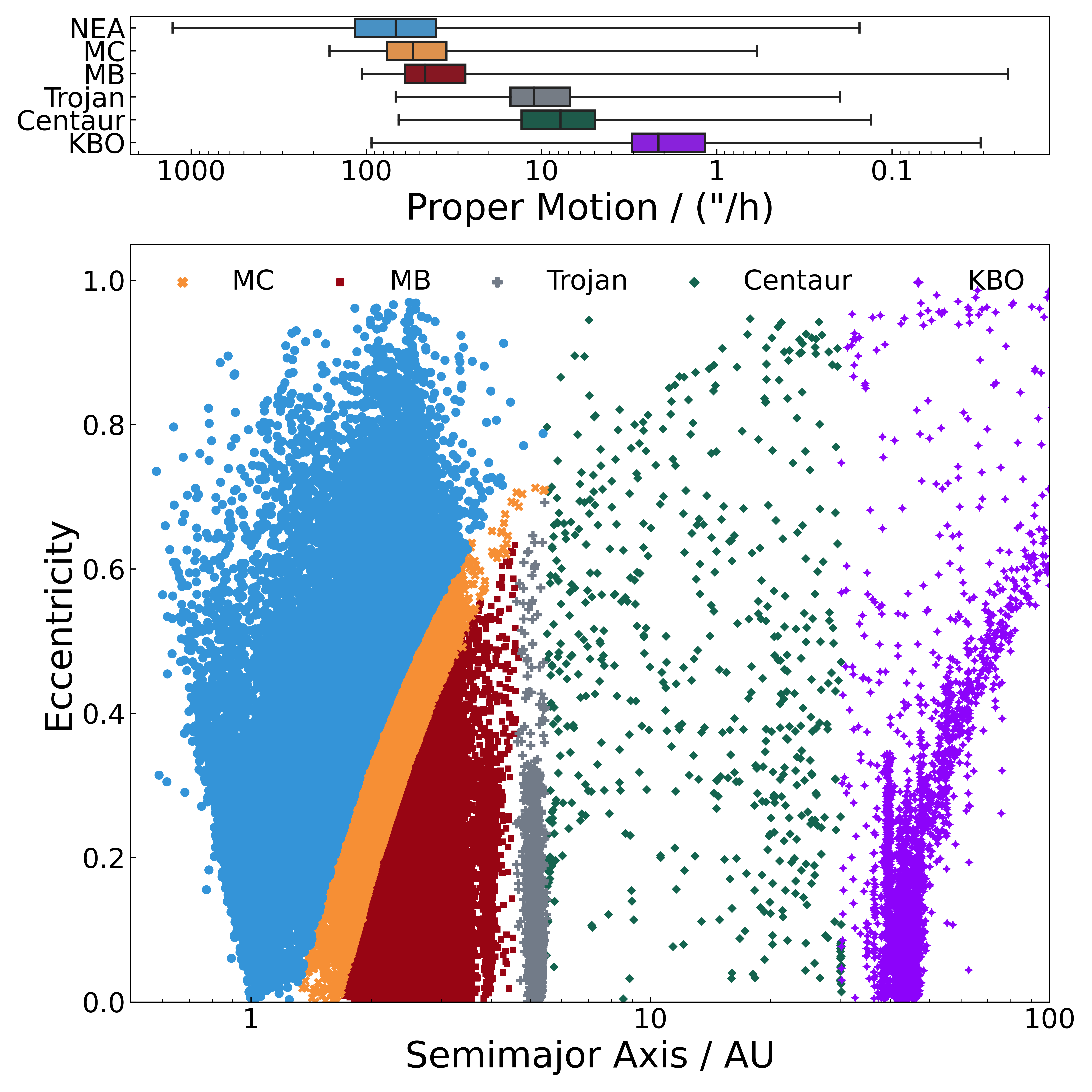}
  \smallskip
  \caption{Distribution of asteroid populations in the Solar System is illustrated based on their semi-major axes and eccentricities from the \astorb database (lower plot). The upper plot shows the distribution of the proper motions for the indicated populations (Near-Earth Asteroids, NEA; Mars-crosser, MC; Main Belt, MB; Trojans; Centaurs; and Kuiper Belt Objects, KBO), extracted at a single epoch for all asteroids. The boxplot displays the minimum and maximum values for each population (whiskers), as well as the 25\,\% and 75\,\% quartiles (box edges) and the median proper motion (box centre).}
  \label{fig:aster_dist}
\end{figure}

The distribution of asteroids in the Solar System is illustrated in Figure\,\ref{fig:aster_dist}, where the bottom part depicts the semi-major axes and eccentricities of all asteroids in the \astorb database. We color-coded the different asteroid populations, which are defined in this orbital parameter space \citep{2008-SSBN-2-Gladman, 2016-Icarus-268-Carry}. Over 90\,\% of asteroids are in the main belt between Mars and Jupiter. The transient population of Near-Earth Asteroids (NEAs) is of great interest for space exploration as these objects represent the celestial bodies closest to Earth. The distant and faint Kuiper Belt Objects (KBOs) are the most challenging to observe, nevertheless, their primordial chemical compositions provide key constraints on models of the formation and evolution of the outer Solar System.

The strategy for the SSOSS cross-match pipeline is based on this distribution, in particular on the distribution of apparent proper motions with respect to each individual satellite point of view, from high-speed NEAs, ($\sim\,$3\,\% of all known asteroids), to relatively slow KBOs ($\sim\,$0.4\,\%).
The upper part in Figure\,\ref{fig:aster_dist} shows the apparent motion distributions of the different populations, which we extracted at a single epoch for all asteroids using the IMCCE SkyBoT3D service\footnote{\url{http://vo.imcce.fr/webservices/skybot3d/}} \citep{2006ASPC..351..367B}. 

The SSOSS input catalogue for comets was provided by the IMCCE \cometpro\footnote{\url{http://www.imcce.fr/en/ephemerides/donnees/comets/index.html}} and it contains the orbital elements of 1,342 comets as of June 20, 2018. For the propagation of positional uncertainties, a fixed initial uncertainty of 10\arcsec was assumed and propagated in time from the proper motion expected values as a first-order approach for the final uncertainty in the position. It is important to note that all serendipitous detections for comets are available in the service, but this is beyond of the scope  of this work, which is focused on the study of the output results for the asteroid sample.

\subsection{Imaging data}

This section describes the selected input imaging data and its associated metadata. For the purpose of this work, only the high-level metadata details describing each data product are required for the cross-match algorithms. These metadata are publicly available through each individual ESA archive, either via direct web access, or via their Table Access Protocol (TAP)\footnote{\url{http://www.ivoa.net/documents/TAP/}} services. This protocol was developed within the scope of the International Virtual Observatory Alliance (IVOA)  and is aimed at providing basic access to all public metadata tables set on each archive or scientific data center.

The standard metadata columns involved in this work are: the start time of the observation (usually referred to as \textit{t\_min} in the IVOA Observation Data Model Core components standard, or ObsCore standard\footnote{\url{http://www.ivoa.net/documents/ObsCore/}}); the end time of the observation (or \textit{t\_max}); the exposure duration (\textit{t\_exptime}); and the instrument footprint or Field of View (FoV) of the observation in sky coordinates (\textit{s\_region}).

From the above time metadata columns, we found that the \textit{t\_exptime} is a more reliable tracker on the total amount of real exposure time than the time difference between the start and the end time of the observation. The difference in time is considered to be the total time required to execute the observation, including time on source, internal calibrations, slewing, settling, etc. In other words, this time difference is usually the effective exposure time on source, plus all the overheads that are required to complete the observation. Hence, the \textit{t\_exptime} metadata was used for the pipeline cross-match software.

However, as shown in Section~\ref{sec:results}, there are some deviations to this definition (i.e., exposure duration values longer than time differences) that can lead to caveats in our catalogue. This is true because, as described in Section~\ref{sec:methods}, this work is particularly sensitive to the time information, so any discrepancies between the published metadata available in the archives with respect to the real timestamp of the observation will introduce false detections in our final catalogue.

Finally, the high-level metadata information available in each individual archive is also integrated and available through the ESASky server, either from the client interface or through the TAP protocol from any Virtual Observatory application \citep[e.g., TOPCAT\footnote{\href{http://www.star.bris.ac.uk/~mbt/topcat/}{http://www.star.bris.ac.uk/~mbt/topcat/}},][]{2005-ASPC-Taylor}. These metadata views are the curated high-level metadata representation for a given observation, that is, it provides a single set of metadata (i.e., row entry) per observation. Thus, these are the final metadata tables used as input in this work.

ESASky integrates all ESA astronomy high-level metadata and associated data products and also provides access to other external facilities (ESO, MAST, and CADC) high-level products via TAP. From the full data set available, three major ESA missions were included in the first version of the SSOSS: HST, \herschel and XMM-Newton. The imaging data used from each mission are described in the following subsections. 

\subsubsection{HST data}
The data used from HST were all available imaging modes and filters from current and past HST instruments, from the start of the first HST scientific imaging observations to the last run of SSOSS (August 17, 1990 -  July 5, 2019). Imaging data were used from the following current HST instruments: the Wide Field Camera 3 (WFC3, 2009-present) in the ultraviolet and optical (UVIS) channel (200 - 1000 nm) and infrared (IR) channel (850 - 1700 nm); the Advanced Camera for Surveys (ACS, 2002-present) in the Wide Field Channel (350 - 1050 nm), High Resolution Channel (200 - 1050 nm) and Solar Blind Channel (115 - 180 nm); and the Space Telescope Imaging Spectrograph (STIS) in imaging modes using the far-ultraviolet Multi-Anode Mirco-channel Array (MAMA; 115 - 170 nm) detector, the near-ultraviolet MAMA detector (165 - 310 nm) and the optical CCD (200 - 1000 nm). Imaging data were also used from the following past instruments: the Faint Object Camera (FOC, 1990-2002) in low, medium and high resolution modes (from 122 to 550 nm); the Near Infrared Camera and Multi-Object Spectrograph (NICMOS, 1997-1999, 2002-2008) in imaging modes using the NIC1, NIC2 and NIC3 channels (800 - 2500 nm); the Wide Field Planetary Camera 2 (WFPC2, 1993-2009), Wide Field Camera images and Planetary Camera images (115 - 1000 nm); and the Wide Field/Planetary Camera (WFPC; 1990-1993), Wide Field Camera images and Planetary Camera images (115 - 1000 nm).

\subsubsection{\herschel\ data}
The ESA {\textsl{Herschel Space Observatory}\xspace} \citep{Her_GLP} was launched in April 2009. For almost four years before running out of coolant, it observed in the far-infrared and sub-millimeter regime with two instruments with photometric capabilities: the Photodetector Array Camera and Spectrometer (PACS) \citep{PACSCalPer} with a three-band imaging photometer (at 70, 100, 160 $\mu$m), and the Spectral and Photometric Imaging Receiver (SPIRE) \citep{SPIRECalPer},  operating a three-band imaging photometer at 250, 350, and 500 $\mu$m. 

The \herschel observing strategy for photometry observations on large areas of the sky was based on the scan technique, as the resulting modulation allowed to reduce the $1/f$ noise of the bolometer readout systems on board. The scan-map technique combines a series of parallel slews together. All the slews (scan legs) must be the same length and the telescope must come to a stop after each scan leg before traversing to the starting point for the next one. The detector array passes over the target area, while taking data continuously along the scan leg. Further improvement was achieved by cross-linked scanning at complementary angles. According to the target visibility window during the observation cycle, observers optimized the sensitivity and sky coverage for their programs by selecting a number of varying parameters within the Astronomical Observation Template (AOT): scan-legs number, length, and separation, along with the scan angle and the scan speed (from 20 to 60\arcsec$/$s). The SPIRE-PACS parallel scan mapping observation AOTs were available to increase the \herschel observatory's efficiency, thus most observations presented here were performed in the SPIRE-PACS Parallel mode ($\sim$62\%), followed by the SPIRE Large Scan mode ($\sim$26\%).  

The SPIRE–PACS parallel scan mode\footnote{More details about the Parallel mode are provided in the \citealt{spire_handbook}.} allowed to obtain photometry simultaneously in two PACS bands (either 70 or 100~$\mu$m and 160~$\mu$m) and all SPIRE bands (250, 350, and 500~$\mu$m). Given the large separation of the PACS photometer and SPIRE photometer footprints (21$\arcmin$), this mode was predominantly used for large maps, for which  scan speeds of 20 and 60\arcsec were offered. In this mode, the orientation angle was fixed to the optimum SPIRE orientation angle (at $+42.4\degr$ and $-42.4\degr$ with respect to the focal plane Z-axis, for scan and cross scan, respectively).  In this mode, the onboard Signal Processing Units averaged 8 consecutive frames, as opposed to the four averaged consecutive frames in PACS prime mode, resulting in a smearing of the PACS PSF,  particularly for the fast scan speed of 60$\arcsec$/s. For the SPIRE large map AOT, users would select between \textit{}{nominal} and "fast"\ scan speeds (30 and 60~\arcsec$/$s) along the lines. In cases where the $1/f $ noise was not a concern, single orient scans were possible to attain a larger sky area coverage in stripes up to 20$\degr$ by 4$\degr$.

The assignment of pointing information to the individual frames by the pipeline assumes a non-moving target. In the case of SSOs, we needed to reassign coordinates to the pixels of each frame by using the calculated position of the target as the reference instead of the centre of the FOV with each new pointing request. Therefore, it was necessary to re-process the (\herschel) images in the object co-moving frame to obtain the appropriate photometry and ascertain the object's position.

\subsubsection{XMM-Newton data}
The {\it XMM-Newton} observatory carries several coaligned X-ray instruments: the European Photon Imaging Camera (EPIC) as well as two Reflection Grating Spectrometers \citep[RGS1 and RGS2,][]{Jansen2001}; the latter were not included in this work. The EPIC cameras consist of two Metal Oxide Semiconductors  \citep[EPIC-MOS,][]{Turner2001} and one pn junction \citep[EPIC-pn,][]{Struder2001} CCD arrays, which have a $\sim$30$\arcmin$ Field of View (FoV) and can offer 5 $-$ 6{\arcsec} spatial resolution and 70 - 80 eV energy resolution in the 0.2 - 10 keV energy band. In addition, {\it XMM-Newton}  has a co-aligned 30-cm diameter optical/UV telescope (Optical Monitor, OM), providing strictly simultaneous observations with the X-ray telescopes \citep{Mason2001}. It has three optical and three UV filters over the wavelength range from 180 to 600 nm, covering a 17 $\times$ 17 arcmin$^{2}$ FoV and a Point Spread Function (PSF) of less than 2{\arcsec} over the full FoV. 

All {\it XMM-Newton} cameras are included in the SSOSS cross-match pipeline and are available through the general service. However, the catalogues included in the context of this work are focused on the study of the particular asteroid population; thus, the analysis and catalogue results only include the OM instrument.

%----------------------------------------------------------------------------------------------------%
\section{Methods}\label{sec:methods}

The software used to compute the list of potential detections was developed with the goal of reducing the cardinality of the geometrical cross-matches needed, thus minimizing the computational cost. This pipeline is based on Java 1.8 and makes use of Java Threads to allow the parallelization of the processes.

In the following subsections, we highlight the main steps from the orbital sampling of sources (Section~\ref{ephemeris_sampling}) to the cardinality reduction of the input list (Section~\ref{cardi_red}) to the geometrical cross-match of each orbital path with the observational footprints (Section~\ref{geom_xmatch}).

\subsection{Ephemerides sampling}\label{ephemeris_sampling}

For this sampling, an even time sampling of the apparent position of the SSO from each satellite point of view is performed first and stored locally for a fixed time interval of ten days. This computation is performed with the Eproc v3.2 suite provided by IMCCE \citep{1998NSTBL..62.....B}. The time span of this sampling is linked with the life-time of each mission. To take into account the satellite reference frame for the ephemerides computation, Eproc software was provided via the SPICE \texttt{spk} kernels\footnote{\url{https://naif.jpl.nasa.gov/naif/data.html}} \citep{2018-ACTON}, with the orbital information for each mission.

In the case of the HST, this file was publicly available at NASA's Navigation and Ancillary Information Facility (NAIF\footnote{\url{http://naif.jpl.nasa.gov/pub/naif/HST/}}), whereas the XMM-Newton kernel was provided by the Science Operations Centre (SOC) at ESAC. Finally, the \herschel Orbital Element Message (OEM) was produced by the SOC and converted in-house at the ESDC into the appropriate SPICE kernel. Table~\ref{table:spicekernels} summarizes the status of the kernels used currently by the cross-match pipeline.

\begin{table}[!ht]
\centering
  \caption[]{Period covered by the SPICE \texttt{spk} kernels of the space missions used in this work.}
    \label{table:spicekernels}
    \small
  \begin{tabular}{lc}
    \hline
    Satellite & Period \\
    \hline\hline
     Hubble Space Telescope & 1990/04/26 - 2019/07/05 \\
    XMM-Newton Space Telescope   & 1999/12/17 - 2019/07/05 \\ 
    \herschel Space Observatory   & 2009/05/16 - 2013/07/01 \\
    \hline
 \end{tabular}
\end{table}

\subsection{Cardinality reduction}\label{cardi_red}
This second step serves as a fast selection of the potential detection of  candidates per mission dataset and reduces the number of geometrical cross-matches needed in the subsequent step, as when it is high, it becomes very costly in terms of CPU times and resources. The position and uncertainty of each object coming from the previous orbit sampling is cross-matched against the selected datasets imaging footprints. 

\begin{figure}[!ht]
  \centering
  \includegraphics[width=.40\textwidth]{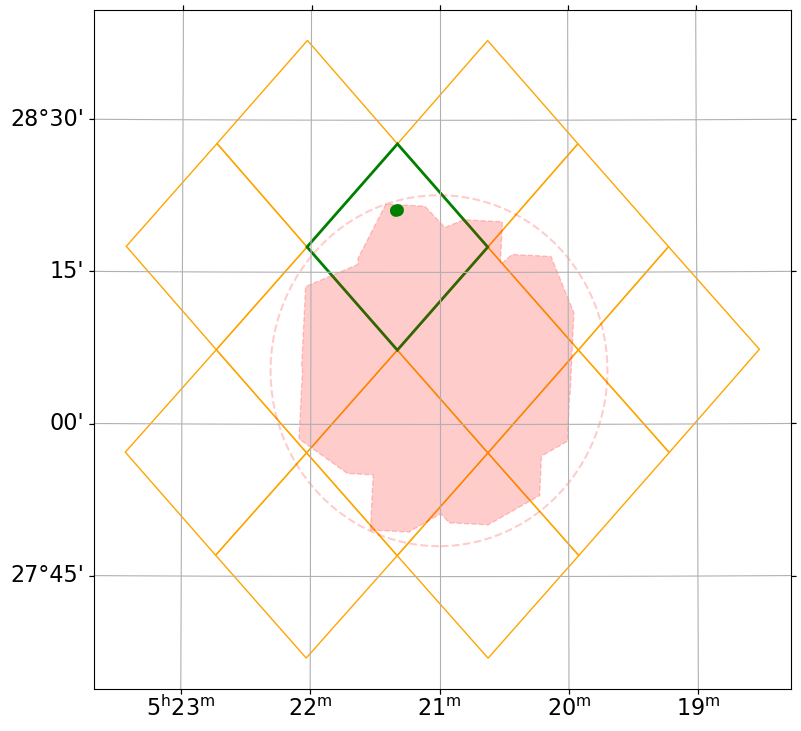}
 \caption{Example of the cardinality reduction step for asteroid \nuna{87}{Sylvia} against \herschel observation id 1342266670, based on the comparison among the HEALPix cells representing the maximum position uncertainty of the asteroid \nuna{87}{Sylvia} over the ten-day period (blue dotted tile), and the HEALpix cells representing the observation footprint in red (orange tiles). The latter are computed as the HEALpix cells overlapping with the smallest circle containing the observation footprint (dotted red circle). The position of Sylvia at the $t_{start}$ of the observation is displayed in blue.
}
  \label{fig:card_reduction}
\end{figure}

\begin{figure}[!ht]
\centering
\begin{tikzpicture}
  [
%    grow                    = right,
    sibling distance        = 6.5em,
    level distance          = 7em,
    edge from parent/.style = {draw, -latex},
    every node/.style       = {font=\footnotesize},
    sloped
  ]

  \node [root] {sso \\ephemerides}
    child { node [env] {rejected}
      edge from parent node [above] {CEU > 1deg} }
    child { node [dummy] {}
      child { node [env] {$d_{max} \propto pm_{max} * \Delta t$}
        child { node [env] {Nside=256}
          edge from parent node [above]
                   {$d_{max} < D_{128}$} }
        child { node [env] {Nside=128}
          edge from parent node [above]
                   {$d_{max} < D_{64}$} }
        child { node [env] {Nside=64}
          edge from parent node [above] {$d_{max} > D_{32}$}}
        child { node [env] {Nside=32}
                edge from parent node [above] {$d_{max} < D_{32}$}}
        edge from parent node [above] {$d_{min}<1AU$} }
      child { node [env] {Nside=32}
              edge from parent node [above, align=center]
                {$d_{min}>1AU$}
              }
              edge from parent node [above] {CEU < 1deg}};

\end{tikzpicture}
 \caption{Decision tree depicting the selection of HEALPix order based on the minimum distance ($d_{min}$) of the SSO to the satellite (first decision level) and on the maximum distance ($d_{max}$) the SSO is expected to cover based on its theoretical proper motion ($pm$) during the time period sampled, where $\Delta t = 10d$ (second decision level). Here, $d_{max}$ is compared to the diameter size for each HEALPix order ($D_{128}$, $D_{64}$, $D_{32}$). An initial pre-selection process of ephemeris based on a CEU threshold of 1 deg is performed.
 \label{fig:healpix_decision_tree}
  }
\end{figure}
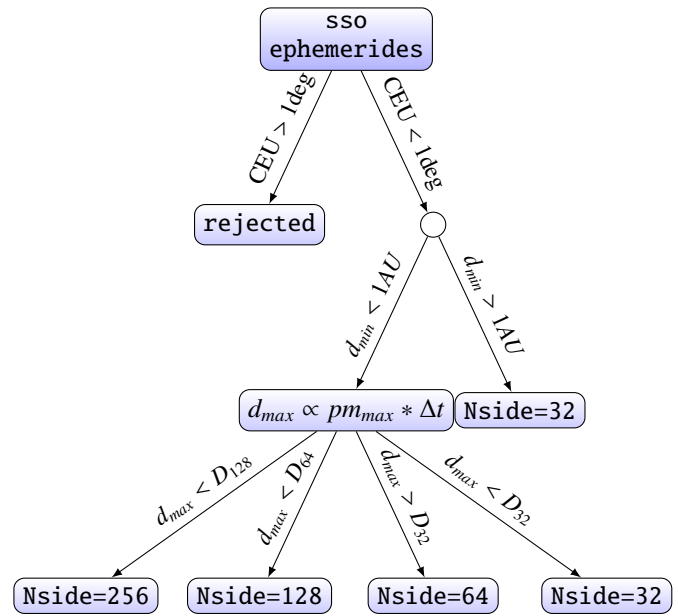

To speed up this selection, this process is based on the HEALPix \citep{2005ApJ...622..759G} tessellation of the sky as illustrated in Fig.\ref{fig:card_reduction}. HEALPix indices are used to represent both the sky-path of each object during each time sample and the observation footprint (representation of the FoV of a given instrument). The selection of the HEALPix \(N_{side}\) is computed based on the minimum distance to the object and its maximum apparent proper motion (Fig.~\ref{fig:healpix_decision_tree}). Furthermore, \(N_{side}\) is the number of pixels per HEALPix side thus proportional to the total number of hierarchical iso-area pixels, so it can be used to trace the minimum pixel area required to fulfill the conditions above.

This is a necessary trade-off between the overall time performance of the pipeline and the completeness of the output list of candidates generated in this step. The more quickly the object moves in the sky, the lower the HEALPix \(N_{side}\) used (i.e., the bigger the pixel areas); thus a greater number of false positives will be included in the output list of potential candidates  and, therefore, more computational time will be required then. In other words, high-proper motion candidates that require HEALPix \(N_{side}\)=32 will degrade the overall performance of the cardinality reduction, but it will ensure there will be no real positives left out in this step.

\subsection{Geometrical cross-match}\label{geom_xmatch}
  
\begin{figure*}[!ht]
    \minipage{0.32\textwidth}
        \includegraphics[width=\linewidth]{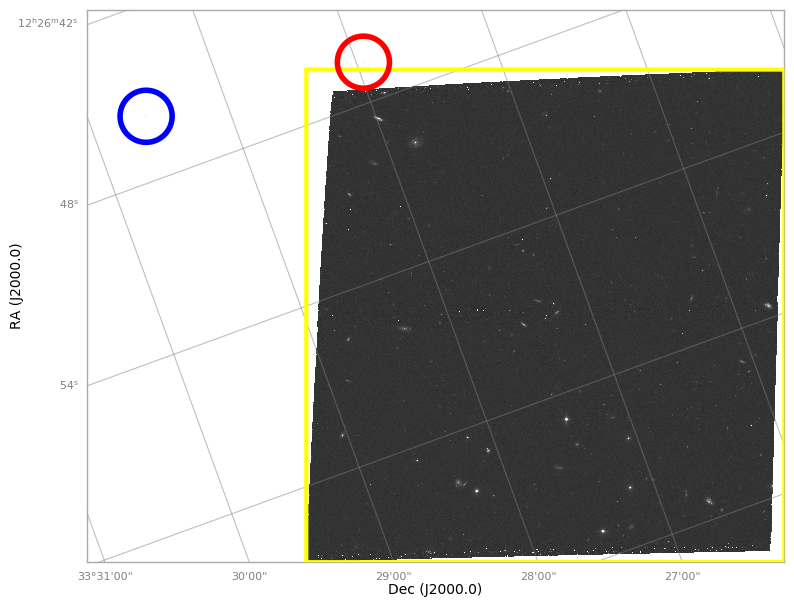}
    \endminipage\hfill
    \minipage{0.32\textwidth}
        \includegraphics[width=\linewidth]{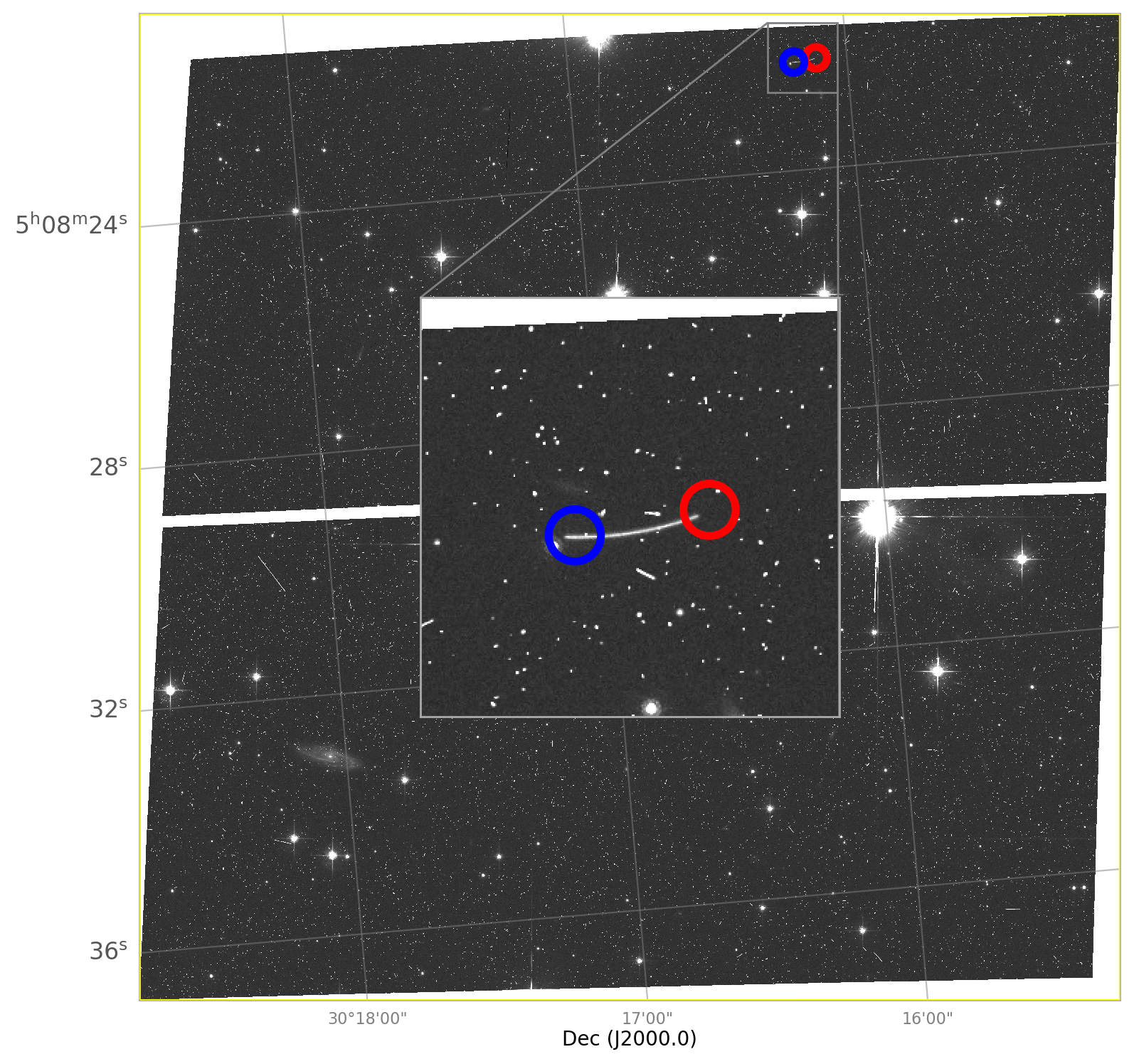}
    \endminipage\hfill
    \minipage{0.32\textwidth}
        \includegraphics[width=\linewidth]{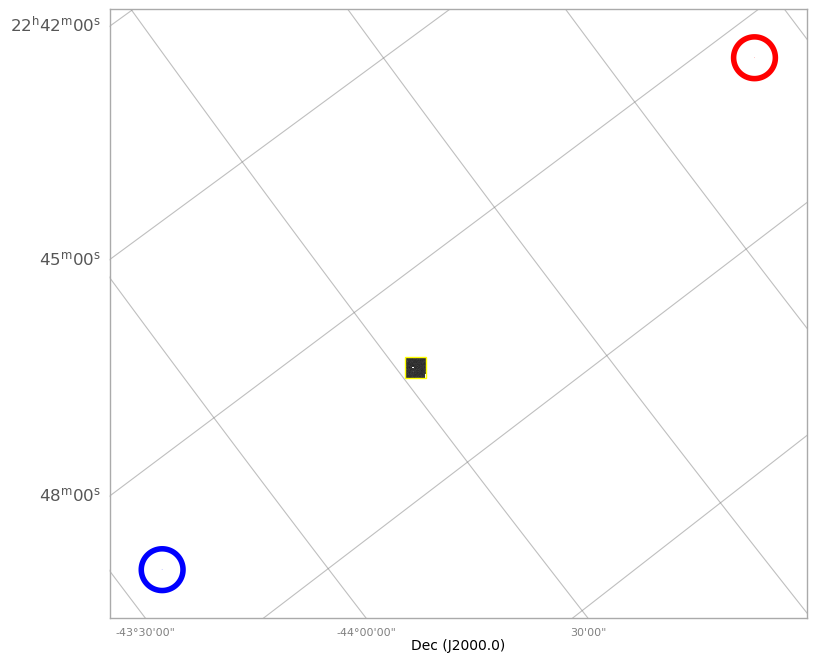}
    \endminipage
    \caption{Examples of the different cross-match types identified in the pipeline. Blue and red circles represent the start and end positions of the asteroid as computed during the geometrical cross-match step. Left: Cross-match type 1, with HST observation j9e106010 and asteroid 2017 MG8. Centre: Cross-match type 2 of HST observation j8g9nnchq and asteroid 2007 TE448. Right: Cross-match type 3 of HST observation jca505010 and asteroid 2014 MW26. }\label{fig:xmatch3in1}
\end{figure*}

The output list of candidate observations per source undergoes then a third step: a new precise geometrical cross-match. At this stage, the position of each object is re-computed to the exact start time and duration of the observation and the cross-match is performed against the observation footprint (FoV).
There are three possible scenarios or cross-match types included  and provided to the final user as illustrated in Fig.\ref{fig:xmatch3in1}: \textit{type 1}: the position of the SSO is not included in the observation footprint, but the uncertainty of the position overlaps with the footprint polygon or contains the footprint polygon; \textit{type 2}: the position of the SSO lies within the observation footprint; \textit{type 3}: none of the SSO predicted positions (start time, end time) nor their uncertainties overlap with a FoV, but the path followed from the start to the end position does cross the FoV of the observation.

This categorization allows for an easy selection of the potential detection candidates for final inspection by the user, namely, the type 2 detection should certainly be included in a particular observation, given the object apparent magnitude remains below the limiting magnitude for the observation, whereas type 1 and type 3 detections are linked to the position uncertainty or to the expected source path overlapping the footprint during the time of exposure; thus the certainty of a real detection is lower. A more detailed description of the algorithms included for each of the cross-match types is included in Appendix \ref{cross-match descriptions}.

The geometrical cross-match is performed by a Java thread taking as input a set of orbital parameters, belonging to a particular source, and the output list of potential cross-matched observations for this particular object, generated by the pipeline itself in the previous step (Section \ref{ephemeris_sampling}). This thread goes sequentially through the list of cross-match algorithms, and as soon as a cross-match type is positive, this process stops and returns the type of cross-match to be ingested later on in the database. The sequential order followed  (type 2 $\rightarrow$ 1 $\rightarrow$ 3 ) is to minimize the algorithm computational time, namely, type 2 is the least time-consuming.

%----------------------------------------------------------------------------------------------------%
\section{Results}\label{sec:results}

The full list of all potential detection candidates from the SSOSS is publicly available via the ESASky TAP server\footnote{\url{https://sky.esa.int/esasky-tap/tap}}, which can be accessed in a variety of ways, for instance, from the VO tool TOPCAT and from the astroquery TAP module \citep{2019-astroquery}.

These tables provide both the start and end time positions and uncertainties for each detection based on the time information in each observation, including the theoretical apparent magnitude in the V band, the theoretical proper motion components, and the distance to the SSO from the satellite's PoV.

In this section, we analyze these results for the asteroid population and describe the different estimations applied to compute the limiting magnitude or flux sensitivity for each of the instruments included in this work to produce a final catalogue of asteroid serendipitous detections above the instrument's sensitivity. This final catalog included in this work is a subset from the tables available via the SSOSS, thus containing the same metadata columns plus an additional column per instrument to ease the identification of potential detections above instrumental noise, either with the theoretical flux value for IR detections (\herschel), the limiting magnitude per observation (XMM-Newton) or the expected signal-to-noise ratio (S/N) per detection (HST). These results are available as a subset of the above-mentioned TAP server, under the \textsl{sso{\_}20190705} schema.

\begin{figure}[!ht]
  \centering
  \includegraphics[width=\columnwidth]{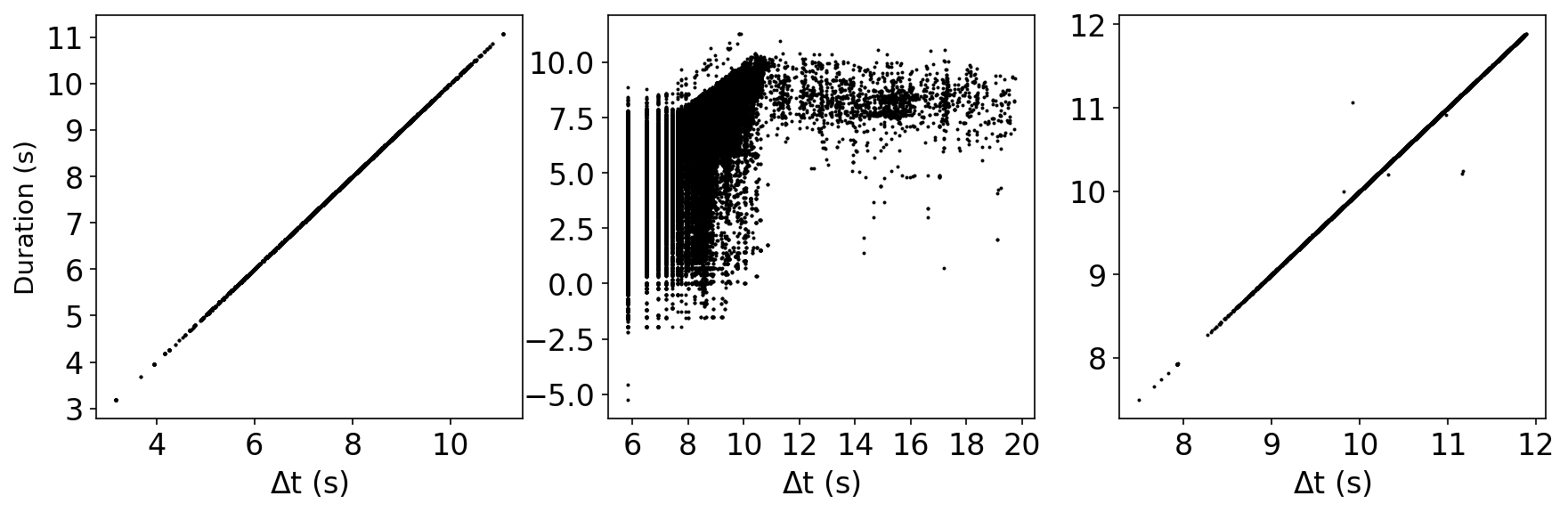}
  \caption{Time interval in seconds between start and end time metadata information versus the exposure duration in logarithmic scale for the three high-level metadata tables involved in this work: \herschel (left panel), HST (center panel), and  XMM-Newton (right panel).}
  \label{fig:timediff_vs_duration}
\end{figure}

To illustrate the output from this service, we provide for each mission a small sample of the brightest serendipitous detections to illustrate the information provided in this catalog (Tables~\ref{table:herschelXmatch_list}, \ref{table:hstlXmatch_list}, and~\ref{table:xmmXmatch_list}). Taking into account the fact that the SSO cross-match pipeline had to be run separately for each mission (different satellite point-of-view), we provide a total of 3 tables of potential detections of asteroids. These tables contain: (1) Object name or preliminary designation as provided by the Lowell Observatory (\astorb), or by the IMCCE (\cometpro); (2) Object Identifier if provided; (3) Observation Identifier; (4) Start Position (J2000 equatorial coordinates RA$_{1}$,Dec$_{1}$ in degrees) of the object from the satellite PoV. This is the predicted position of the object at the start time of the observation, provided by the observation metadata stored in the ESDC astronomy archives; (5) End Position (\textsl{idem}, RA$_{2}$,Dec$_{2}$ in degrees) of the source from the satellite point-of-view at the end time of the observation, computed as the metadata start time plus the exposure duration (see Fig.~\ref{fig:timediff_vs_duration} for more details); (6) Start and end position uncertainties in degrees, derived from the current 1-$\sigma$ ephemeris uncertainty (CEU) in arcsec and the rate of change of CEU in arcsec/day. Thus the uncertainty of a predicted ephemeris is derived as $\delta (\alpha,\delta) = ceu + \dot{ceu} \lvert{\Delta t}\rvert$; (7) Predicted apparent proper motion at the start ($\mu_{RA}\cos{(dec)}_{1}$,$\mu_{dec1}$) and end time of the observation ($\mu_{RA}\cos{(dec)}_{2}$, $\mu_{dec2}$) in arcsec/min; (8) Apparent magnitudes in V band ($m_{v1}$,$m_{v2}$) at the start and end of the observation; (9) Distance (au) from the satellite to the object ($d_{1}$,$d_{2}$); (10) Phase angle (deg) as the angle between the Sun and the satellite as seen from the object($phase_{1}$,$phase_{2}$); (11) Elongation (deg) as the angle between the Sun and the object as seen by the satellite point-of-view ($elong_{1}$,$elong_{2}$); (12) Cross-match type ($xtype$), as described in the previous section (Section \ref{geom_xmatch}).

The total number of potential geometrical cross-match detections, targeted versus the serendipitous detections per satellite is presented in Table \ref{table:summary_dets}. The current implementation of this service provides the entire list of geometrical cross-matches between SSOs trajectories and instrument FoV. Many objects may however be too faint to be detected in each observation, depending on the specific limiting magnitude or flux sensitivity integration time and energy band of each observation. Thus, in the following subsections, we analyze each instrument individually to present a selection of candidates above instrumental threshold. Future releases of the SSOSS will include this computation of the apparent magnitude of the SSOs at the wavelength of the observation automatically.

\begin{table}
\centering
\small
  \caption[]{Summary table with total number of potential asteroid detections (i.e. number of distinct observation/asteroid position pair) ($N_{total}$), number of distinct objects (N$_{sso}$), number of targeted observations (N$_{target}$) and number of serendipitous detections theoretically above instrumental limiting magnitude or flux (N$_{>limit}$.}
    \label{table:summary_dets}
  \begin{tabular}{lllll}
    \hline \hline
   Mission &N$_{total}$ & N$_{sso}$ & N$_{target}$ & N$_{>limit}$ \\
    \hline
    \herschel  & 337328  & 114746 & 2010 & 3492 \\ 
    HST        & 134,974 & 14,367 & $\sim$ 80,000 & $\sim$ 32,000 \\ 
    XMM-Newton (OM) & 25138  & 21613  & 0 & 985 \\
    \hline
 \end{tabular}
\end{table}

Finally, it is important to note that the SSOSS was developed with the aim at providing a mission-agnostic tool to allow SSO detection discoveries within the existing ESA astronomy archives. The standardization of this service does not allow full flexibility to accommodate the individual particularities of each mission data set. This is particularly true for the time-related information, since this service is very sensitive to it, providing the exact position at the theoretical start and at the end of the observation. An example of these deviations can be found in Fig.~\ref{fig:timediff_vs_duration}, where the delta time between the start and the end time versus the reported duration are plotted for each data set. 

\subsection{\herschel}
Among the 27,105 \herschel public imaging observations, 2269 specifically targeted SSOs, identified by the naif\_id metadata associated to each observation in the archive. Once the pointed observations of major Solar System bodies (naif\_id < 1000) are removed, 2004 targeted observations remain, plus 6 extra observations not categorized with naif\_id but including major Solar System bodies in their target name (i.e., ``mars offboresight''), representing 7.4\% of \herschel imaging time devoted to observations of asteroids and comets. A large fraction of these targeted observations ($\sim$30$\%$) are included within the "TNOs are cool" program \citep{2009TNOs} targeting KBOs (naif\_id > 2100000). Once removed, the total number of targeted asteroids is 87. However, the total number of potential serendipitous observations is greater by three orders of magnitude.

\begin{figure}[!htb]
  \centering
  \includegraphics[width=.5\textwidth]{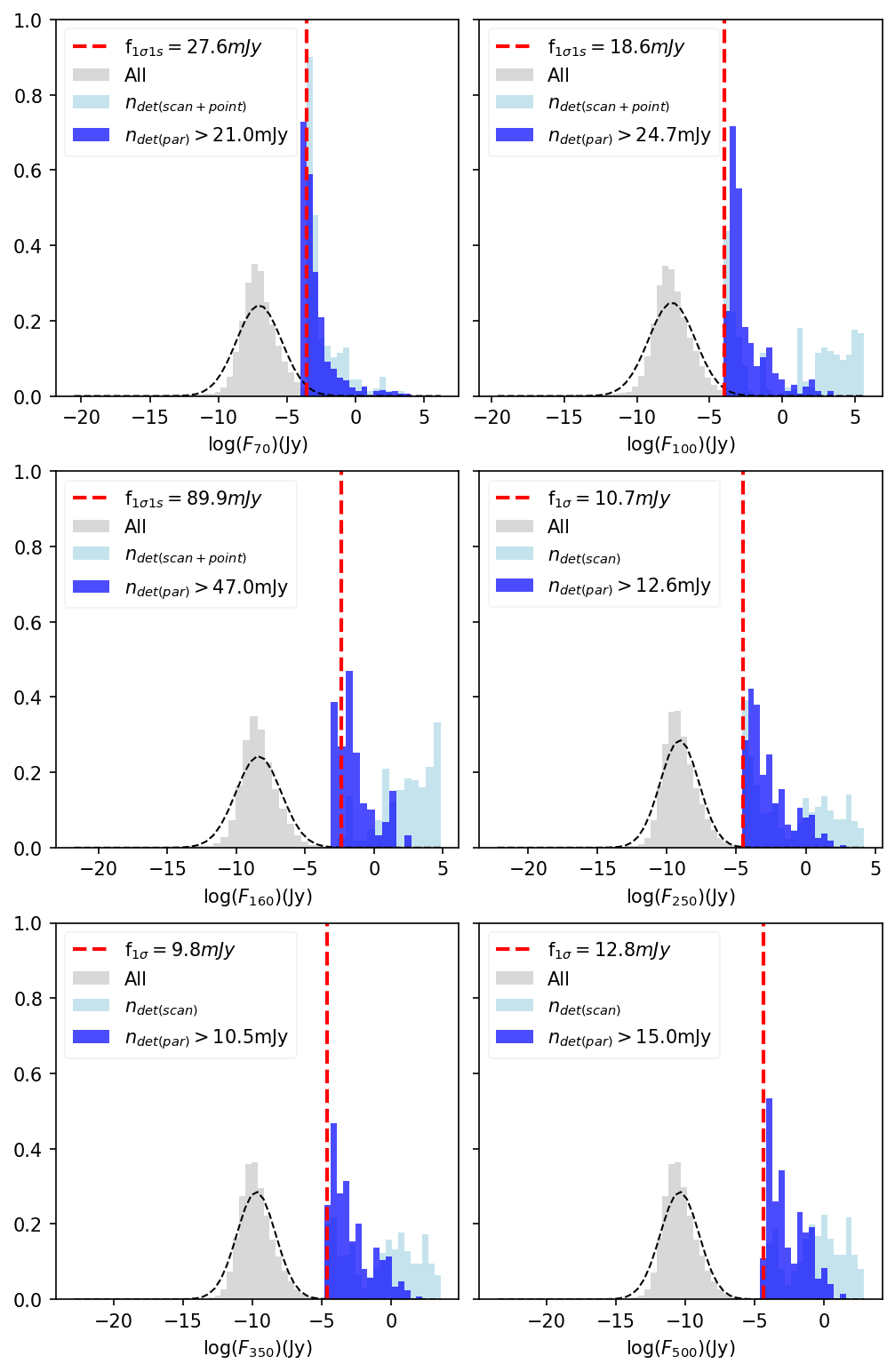}
  \caption{Theoretical infrared flux distribution of geometrical cross-matches for all \herschel bands and fixed input albedo, beaming and emissivity parameters. From top to bottom and left to right: 70, 100, 160, 250, 350, and 500 $\mu$m normalized flux distribution in grey. Red-dashed line marks the estimated flux sensitivity limit per band. The subset detections above flux sensitivity are displayed in blue.}
  \label{fig:Herschel Flux Sensitivity}
\end{figure}

In order to get a rough estimate on the real serendipitous detections above the \herschel flux sensitivity, the theoretical flux value per asteroid and observation pair is calculated at each of the \herschel bands: PACS at 70 or 100 and 160 $\mu$m, as well as SPIRE at 250, 350, and 500 $\mu$m. For this first-order approach, the albedo, emissivity, and beaming parameters were averaged over the set of published values collected from literature when available \citep{2010AJ....140..933R,2011ApJ...741...68M,2020PSJ.....1....5M,2011ApJ...743..156M,2011ApJ...742...40G, 2012ApJ...744..197G,2013ApJ...773...22B,2013ApJ...762...56U,2015ApJ...814..117N, 2016AJ....152...63N, 2018AA...612A..85A},
and a standard set of values across the entire set of potential detections of albedo=0.15, emissivity=0.9, and beaming=1.0 otherwise. These values are well-suited for main-belt asteroids and NEAs, although they deviate from expected parameters across the entire population of asteroid families (in particular KBOs). 

\begin{figure}[!ht]
  \centering
  \includegraphics[width=.50\textwidth]{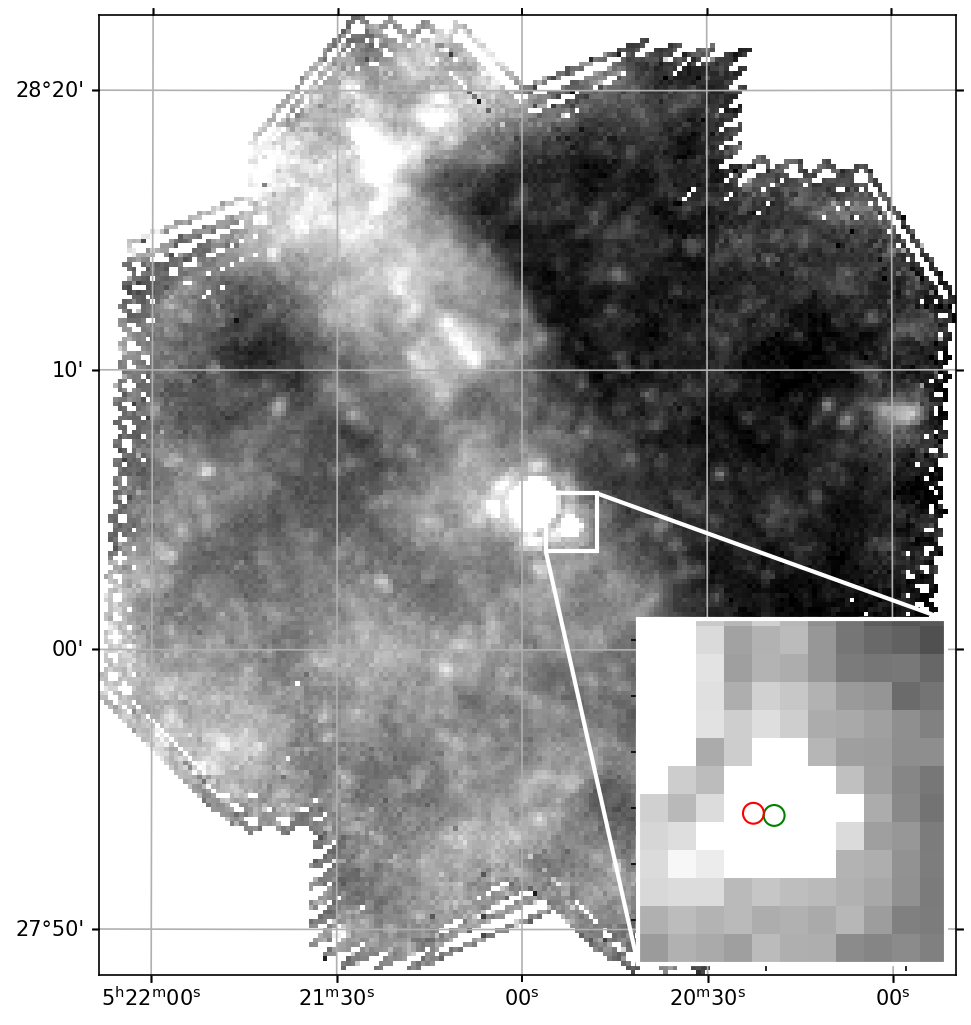}
  \caption{Serendipitous detection of asteroid \nuna{87}{Sylvia} in the \herschel observation 1342266670 targeting \nuna{1}{Ceres} (at the centre of the image) with the PACS instrument on 2013-03-01 03:26:26.0. Both predicted coordinates at the start time (green) and end time (red) lie within the instrument FoV (see zoomed insert).}
  \label{fig:sylvia_in_ceres}
\end{figure}

Based on the estimated photometer sensitivities included in Table 3.5 \citep[from the][]{spire_handbook}, the $1\sigma$ instrument noise for PACS-SPIRE parallel mode for one repetition, nominal scan direction, and scan speed of $60\arcsec/s$ (highest noise values) are 21.0, 24.7 and 47 mJy for PACS at 70, 100, and 160$\mu$m bands, and 12.6, 10.5, and 15 mJy/beam for SPIRE at 250, 350, and 500$\mu$m. 

In the case of SPIRE-only scan mode, we calculated the combined estimated noise levels reported in Table 3.4 of the same document, where the $1\sigma$ instrument noise levels for the nominal scan speed ($30\arcsec/s$) at 250, 350 and 500 $\mu$m bands are 9.0, 7.5, and 10.8 mJy, and the 1$\sigma$ extragalactic confusion noise is 5.8, 6.3, and 6.8 mJy/beam, respectively. Finally, for the remaining observations in PACS-only mode, we included the $1\sigma$,1s integration time (27.6, 18.6, and 89.9 mJy respectively for 70, 100, and 160 $\mu$m) reported in Table 4.3 from the \citep{pacs_handbook}.  These flux sensitivity values, based on the observing mode, are included in Fig.\ref{fig:Herschel Flux Sensitivity}.  

The total number of real potential detections above the \herschel instrumental flux sensitivity are 2546 for the PACS instrument (656 serendipitous detections once targeted observations are removed) and 946 (253 serendipitous) detections in the case of SPIRE, representing only around 0.8$\%$ of the total number of detections available through the SSOSS TAP service under the schema \textsl{sso\_20190705}, and table name \textsl{xmatch\_herschel\_aster}. This table provides an extra boolean column (\textsl{is\_visible}) to show which entries are above the theoretical flux limit.  It is important to note that for the \herschel observatory, this service provides about 25\% more asteroid observations than scheduled. An example of one of these detections can be found in Fig.\ref{fig:sylvia_in_ceres}, where asteroid (87) Sylvia was serendipitously observed in a targeted observation on (1) Ceres \citep{2014Natur.505..525K}.

These rough flux estimates provides a first order-of-magnitude on the number of detections present in the archive. However, for a specific object, it would be advisable to check all observations that are not much fainter than the expected limiting magnitude included in this work.  Currently, a study is underway to extract the PACS fluxes of all serendipitously seen asteroids, combined with a detailed radiometric study (Szakats et al., in preparation).

\begin{table}[!ht]
\centering
  \caption[]{Summary with total number of potential detections for \herschel PACS and SPIRE instruments per observing mode. Here, N$_{>limit}$ is the number of potential detections above the threshold flux sensitivity per instrument mode. This number includes those potential detections having at least one flux value above the given limit (i.e., an observation with two filters, it is included when the theoretical flux of the object is above any of those filters). Finally, N$_{serend}$ represents the serendipitous detections included in N$_{>limit}$, once the targeted sample is removed.}
  
    \label{table:herschel_obs}
    \small
  \begin{tabular}{lllll}
    \hline
   Instrument & Observing Mode & N$_{total}$ & N$_{>limit}$ &N$_{serend}$\\
    \hline
    SPIRE  & SpirePacsParallel   & 134111 & 365 & 2 \\
    SPIRE  & SpirePhotoLargeScan & 114717  & 355 & 44 \\ 
    SPIRE  & SpirePhotoSmallScan & 10651  & 226 & 207\\
    PACS   & SpirePacsParallel   & 134034 & 1658 & 3  \\  
    PACS   & PacsPhoto           & 38126  & 888 & 653\\
    \hline
 \end{tabular}
\end{table}

\subsection{XMM-Newton}
XMM-Newton output lists with potential detections are divided by instrument. This is due to the intrinsic nature of our object sample, since there are not yet any known asteroid detections in the X-ray energy regime, the X-ray EPIC cameras is not included for the catalogue produced in the context of this work. However, we included these geometrical cross-matches in the general SSOSS.

\begin{figure}[!ht]
  \centering
  \includegraphics[width=.5\textwidth]{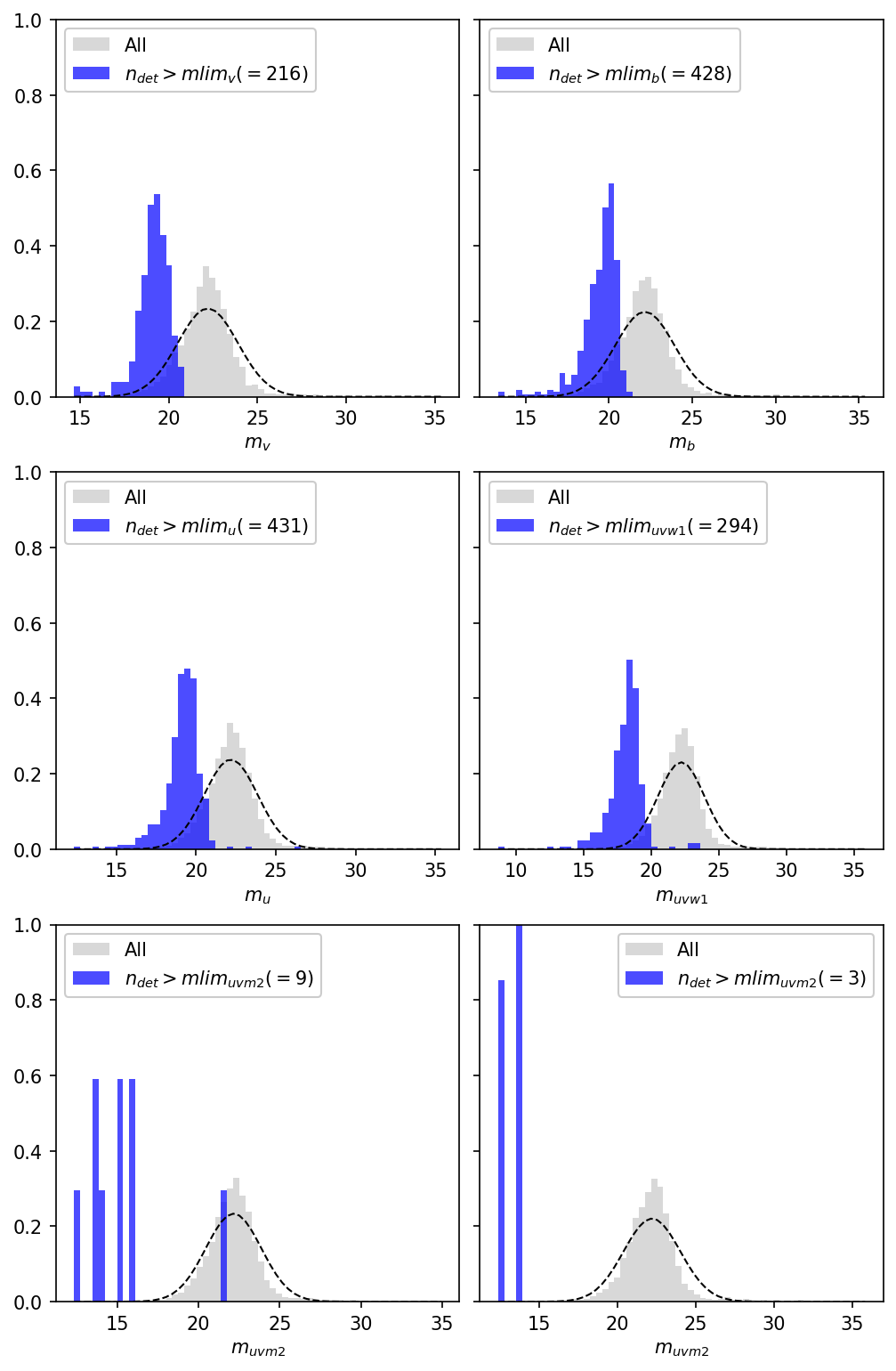}
  \caption{Theoretical apparent magnitude distributions for each OM instrument filters in gray. The subset list of detections above limiting magnitude per observation are displayed in blue. From top to bottom and left to right: V, B, U, UVW1, UVM2, UVW2. Dashed lines represent mean and $\sigma$ of the normal distribution per band.}
  \label{fig:OM mag distribution}
\end{figure}

The summary of the Optical Monitor (OM) potential detections is included in Table~\ref{table:xmm_obs_aster_summary} and Fig.~\ref{fig:OM mag distribution}. Limiting magnitudes per observation id and filter are extracted from the OM pipeline products retrieved via the archive astroquery python module\citep{2019-astroquery}. The estimated limiting magnitudes are provided as header keywords (MLIM<filter>) in the per-observation combined source-list files (*OBSMLI*) where <filter>=V,B,U,W1,M2,W2. These are the 5$\sigma$ field estimates based on the mean background rate, so they are not accurate for specific regions of the image, as the background level can vary substantially and rapidly across the image, but they provide a good first-order estimate of the limiting magnitude.

For the estimation of the potential detections above the instrumental threshold, the output theoretical V magnitude provided by the pipeline needs to be converted for each band to account for the different filter transmission and intrinsic spectral energy distribution of SSO.
We compute these color corrections (listed in Table~\ref{table:xmm_obs_aster_summary})
by retrieving the filter transmission for the SVO filter service
\citep{2012-IVOA-Rodrigo} and assuming SSOs have the same spectrum as the Sun
\citep[which we take from][]{2004-SoEn-76-Gueymard}.

\begin{table}
  \caption{Summary with the total number of asteroid detections for the XMM-Newton OM camera per filter. N$_{serend}$ is the number of serendipitous potential detections above the OM camera limiting magnitudes per filter, provided by the instrument pipeline source list product per observation\_id.}
    \label{table:xmm_obs_aster_summary}
    \small
    \centering 
  \begin{tabular}{l l l l}
    \hline
    \hline
   Instrument & Zero-Point\tablefootmark{a} & N$_{total}$ & N$_{serend}$ \\
    \hline
    OM V      & $-0.0474$ & 4038  & 216\\
    OM B      & $-0.6028$ & 3939  & 428\\ 
    OM U      & $-0.7439$ & 7792  & 431\\
    OM UVW1   & $-1.4842$ & 16501 & 294\\  
    OM UVM2   & $-4.0778$ & 13112 & 9\\
    OM UVW2   & $-3.6144$ & 6605  & 3 \\
    \hline
 \end{tabular}
 \tablefoot{ 
 \tablefoottext{a}{Color correction zero point values from the SVO Filter Profile Service\footnote{\url{http://svo2.cab.inta-csic.es/theory/fps/}}
 \citep{2020sea..confE.182R, 2012-IVOA-Rodrigo}}
 }
\end{table}

It should be noted that the limiting magnitudes provided by the OM pipeline are integrated over the total exposure duration of the observations, which is not necessarily the case for our objects due to the apparent proper motion of our samples. Therefore, the effective exposure duration can be lower and, hence, these limiting magnitudes should be interpreted as the upper limits. 

\begin{figure*}[!ht]
 \begin{minipage}[l]{\linewidth}%{2\columnwidth}
  \includegraphics[width=\linewidth]{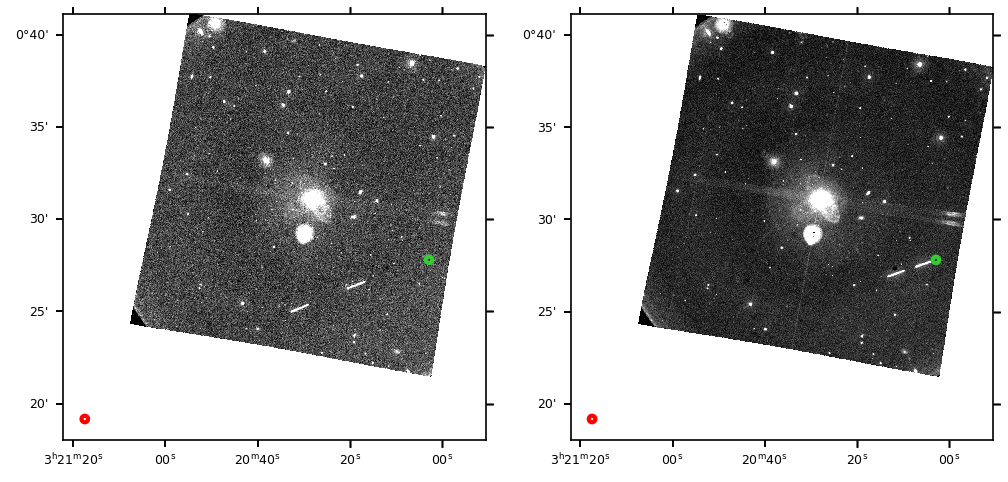}
  \caption{Serendipitous detection of asteroid (234) Barbara in XMM-Newton observation 078104010 on 2016-08-27 22:03:30, captured by the OM camera, at the time observing sequentially with 4 out of 7 available filters. The object remains in the FoV for the entire duration of the observation, only visible at U (left) and UVW1 (right) filters (344nm and 291nm effective wavelength, respectively). The green and red circles represent the position of Barbara at the start time and stop time of the observation, respectively.}
  \label{fig:barbara_om}
  \end{minipage}
\end{figure*}
\smallskip

The total number of theoretical detections grouping the results per observation and sso pair (independently of the number of filters per observation) is 985, representing $~3.9\%$ from the total number of potential detections (25138). An example of these results is displayed in Fig. \ref{fig:barbara_om}, with the serendipitous detection of (234) Barbara in two of the OM filters.

%------------------------------------------------------
\subsection{HST}

% - HST estimations by MM
For HST, the total number of detections returned by the cross-match is 165,607. After removing planets and natural satellites from the sample, 134,974 observations remain of 14,367 distinct asteroids. Using the user-set \texttt{TARGET} and \texttt{MOVING\_TARGET} metadata in the eHST archive, we identified about 58\,\% of the computed observations to be targeted at these SSOs. For the remaining approximately 55,000 serendipitous observations, we modeled the source spectra in a first-order approximation using the STIS solar spectrum \citep{Bohlin_2014}. The solar spectrum was scaled to match the predicted source magnitude in the respective HST observation band. This magnitude was computed using the predicted V-band magnitudes provided by the cross-match and color-correction terms derived from the solar spectrum for all pairs of the V-band and the HST filters. We then computed the expected HST count rates using the \texttt{pysynphot} \texttt{python} package by the STScI, as well as an estimated background rate, yielding an approximate signal-to-noise ratio for each asteroid depending on the exposure time, observation wavelength, and predicted source magnitude.

Simulating the observations revealed that about 59\,\% ($\sim$~32,000) of the serendipitous detections should exhibit signal-to-noise ratios  above 3 and could thus be identified in the images. However, the actual number of serendipitous observations will be lower due to the ephemerides uncertainty of the SSOs: only 6\,\% of the potential serendipitous observations belong to known SSOs where the orbit uncertainty is below 202\,", about the size of one edge of the FoV of the Advanced Camera for Surveys (ACS) aboard HST. It is important to note that  due to the inconsistencies in the reported start and ending times of the observations in the published metadata, a fraction of the cross-matches will be false detections.

\subsubsection*{Zooniverse citizen science project}
\begin{figure}[!ht]
  \centering
  \includegraphics[width=.40\textwidth]{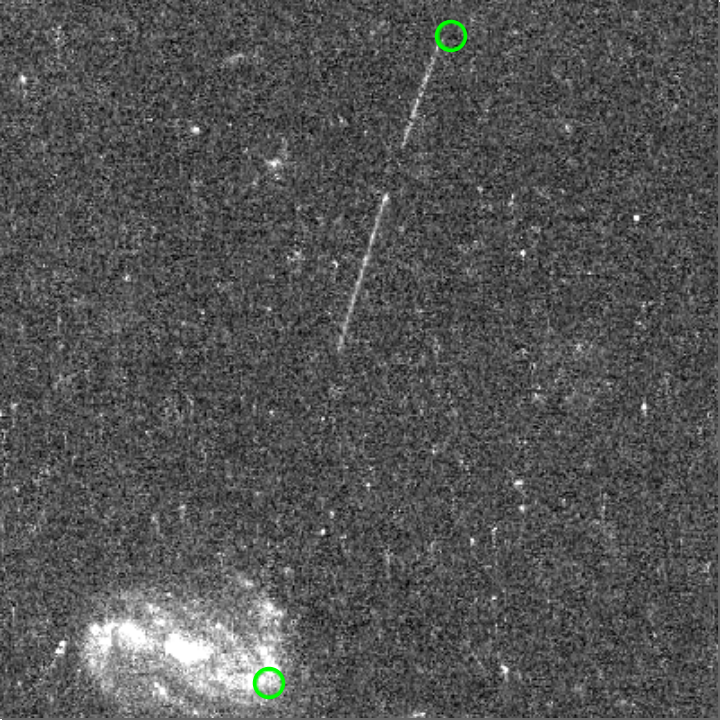}
  \caption{Serendipitous observation of asteroid 2000\,NH10 by the ACS on board HST is visible towards the upper part of the image. The green markers indicate the predicted start- and end-position of the asteroid as computed by the pipeline. The observation itself targeted the galaxy cluster MACS1115+0129.}
  \label{fig:hst_example}
\end{figure}

A first practical application of the SSOSS was demonstrated in the \textit{Hubble Asteroid Hunter}\footnote{\url{https://www.zooniverse.org/projects/sandorkruk/hubble-asteroid-hunter}} citizen science project on the Zooniverse platform. We extracted cutout images of potential serendipitous detections where the asteroid was predicted to be in the frame at the start- or end-epochs of the HST observation, or to cross the FoV in the meantime (types 2 and 3), to reduce the number of false positives introduced by the ephemerides uncertainty of the asteroids. The volunteers had to mark the actual position of the asteroid trails in the cutouts. An example cutout image is shown in Fig.\,\ref{fig:hst_example}, showing part of an HST/ACS observation of the galaxy cluster MACS1115+0129. The image consists of two combined exposures, each containing the trail of asteroid 2000\,NH10 towards the upper part of the image. There is a gap visible in the asteroid trail due to the observation gap between the two exposures. 

We further extended the project to search for serendipitous observations of unknown asteroids or asteroids with large ephemerides uncertainties by providing cutouts of all the HST ACS and WFC3 observations to the volunteers. In total, 11,000 volunteers inspected over 150,000 images in the Hubble Asteroid Hunter project during a period of one year. The results from this project will be presented in a forthcoming paper (Kruk et al., in prep.).

%----------------------------------------------------------------------------------------------------%
\section{Thermophysical analysis of (16) Psyche}\label{sec:psyche}

We illustrate a use case of the SSOSS service by analyzing the mid-infrared fluxes of the  large main-belt asteroid \nuna{16}{Psyche}, serendipitously observed by \herschel\ and\ the target of the NASA \textsl{Psyche} mission \citep{Elkins2017}. Historically, \nuna{16}{Psyche} has been the archetypal metal asteroid. However, based on estimates of its bulk density \citep{Viikinkoski2018, 2020AA...638L..15F}, its visible and near-IR spectrum
\citep{Landsman2018}, and the variation of its radar albedo over its surface \citep{Shepard2015},
it is now considered to be more likely a mixture of metal and silicates \citep[see for a review][]{Elkins2020}. Its metal content, composition, and meteoritic analog(s) are still highly uncertain.

One of the diagnostic parameters for a metallic versus silicate composition is the thermal inertia of a body. Thermal inertia, defined as $\Gamma = \sqrt{k\rho c}$, with the thermal conductivity, $k$, the density, $\rho$, and the specific heat capacity, $c$, is a measure of how quickly the surface temperature of an object adapts to changing solar energy input. The thermal inertia of large silicate asteroids are low, generally below 100 J\,m$^{-2}$\,s$^{-1/2}$\,K$^{-1}$ \citep[e.g.,][]{AliLagoa2020}, as those asteroids are covered by a thermally isolating, powdery regolith layer. The thermal inertia of asteroids containing a larger fraction of metal is expected to be higher, due to the larger conductivity.

There are two determinations of the thermal inertia of \nuna{16}{Psyche}: \citet{Matter2013} find a thermal inertia of 243-284 J\,m$^{-2}$\,s$^{-1/2}$\,K$^{-1}$, consistent with a metal-rich object. On the other hand, \citet{Landsman2018} arrive at a much lower thermal inertia of 11-53 J\,m$^{-2}$\,s$^{-1/2}$\,K$^{-1}$, which is more consistent with a silicate composition. Here, we add far-infrared observations to the existing data-sets of thermal properties of \nuna{16}{Psyche}, and investigate whether those additional data allow us to remove the ambiguity about \nuna{16}{Psyche}'s thermal inertia.           

\begin{figure}[!ht]
  \centering
  \includegraphics[width=.50\textwidth]{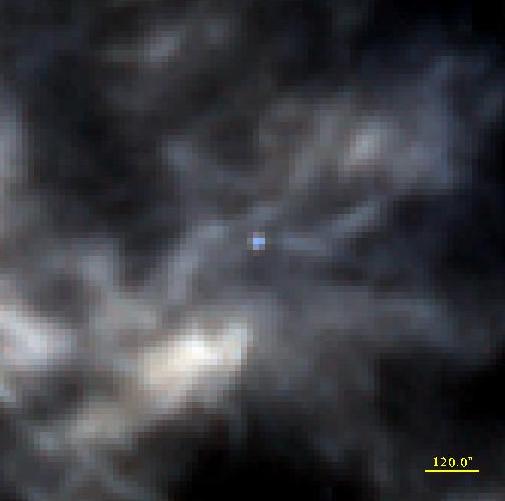}
  \smallskip
  \caption{Serendipitous detection of asteroid Psyche 16 in \herschel observation 1342202250 with the SPIRE instrument. Pseudo-color image with 250 $\mu$m in blue, 350 $\mu$m in green, and 500 $\mu$m in red. The asteroid is in the centre, the background is Galactic dust emission. North is up, east to the left.}
  \label{fig:psyche_16}
\end{figure}

We introduced in this work the first observations of this asteroid in the far-infrared regime (70- 500 $\mu$m), thanks to the two serendipitous observations made by \herschel in 2010 (\texttt{OBS\_IDs} 1342202250, 1342202251, which target the Galactic region L1551). These are included in Table~\ref{table:herschelXmatch_list} as seen in Fig.~\ref{fig:psyche_16}. 

Both observations are in parallel mode, where PACS and SPIRE instruments observe simultaneously. Observation 1342202251 is at nominal scan direction, while 1342202250 is at an orthogonal direction, both at a fast scan speed (60\arcsec/s). The detector footprints of both instruments passes over the target for a very short time (seconds) and, hence, it is not necessary to convert the timelines and the images to the asteroid reference frame. The times for each detection of Psyche in PACS and SPIRE are listed in Table~\ref{table:herschel_psyche}. As the times in each observation are separated by $\sim$1h23 minutes, combining the two images without converting to the SSO reference frame will result in a double source, as it appears in ESASky in the PACS and SPIRE HiPS.

\begin{table*}[!ht]
  \caption[]{Summary of the \herschel photometry for Psyche. The total flux densities at 70, 160, 250, 350, and 500 $\mu$m in Janskys. The epochs of the peak flux detection of the source when it was scanned by the detectors are provided in the column marked ``Epoch.'' We note that the PACS and SPIRE scans in Parallel mode are separated by $\sim20\arcmin$, and that is why the asteroid is not seen at the same time, but with an 8 minute offset.}
    \label{table:herschel_psyche}
    \small
    \centering
  \begin{tabular}{lllccccc}
    \hline
   Instrument & \texttt{OBS\_ID} & Epoch &  \multicolumn{5}{c}{Photometry (Jy)} \\
                                         &  &  & F70 & F160 &  F250 & F350 & F500 \\
\hline
 PACS & 1342202250 & 2010-08-07T17:30:40 & 19.49 $\pm 1.00$ & 4.58 $\pm 0.28$ & & & \\
 PACS & 1342202251 & 2010-08-07T18:54:57 & 18.98 $\pm$ 0.98 & 4.59 $\pm$ 0.34 & & & \\
\hline 
 SPIRE & 1342202250 & 2010-08-07T17:22:05 &  & & 1.47 $\pm$ 0.10 & 0.61 $\pm$ 0.05 & 0.42 $\pm$ 0.03 \\
 SPIRE & 1342202251 & 2010-08-07T18:46:39 &  & & 1.63 $\pm$ 0.10 & 0.90 $\pm$ 0.06 & 0.61 $\pm$ 0.04 \\
\hline
 \end{tabular}
\end{table*}

We extracted the flux densities of Psyche in each observation independently. We assumed the source is point-like in all \herschel bands. For PACS, we performed aperture photometry and corrected it for aperture and color of the source. The flux errors include the aperture flux error, the flux calibration uncertainty of 5\%, and 1\% color correction uncertainty, all added in quadrature. For SPIRE, we used the Sussextractor \citep{Suss} source extraction method and obtained the total flux density in the three SPIRE bands. The fluxes are corrected for color, assuming the source has power-law spectrum with index 2 (i.e., blackbody spectrum in Rayleigh-Jeans regime) and we also applied a pixelisation correction (see \citealt{spire_handbook}). The SPIRE flux errors are the photometry error, the flux calibration uncertainty of 5.5\%, and 1.5\% uncertainty on the pixelisation correction, all added in quadrature. In fast scan Parallel mode observations, the instrumental noise is the dominant source of error and it is already incorporated in the flux error estimate from photometry.

The PACS and SPIRE measurements took place on Aug 7, 2010 between 17:22 and 18:55 UT when Psyche was 2.56\,au from the Sun and 2.77\,au
from \herschel, seen under a phase angle of 21.7\degr\space and an aspect angle of 148.4\degr\space (see Figure~\ref{fig:PsycheShapeTPM}).
The high aspect angles means that \herschel has predominantly seen  the object's southern hemisphere
(an aspect angle of 180\degr\space would indicate a perfect south pole view), where the shape model
presents two crater-like depressions \citep{2017-Icarus-281-Shepard,Viikinkoski2018}.

For our radiometric study, we applied the thermophysical model 
\citep[described in][]{Lagerros1996, Lagerros1997, Lagerros1998, Muller1998, Muller-Lagerros-2002}.
We used the \citet{Viikinkoski2018} spin-shape solution (discarding its absolute size information) for the interpretation of our extracted PACS and SPIRE fluxes. 

Psyche's H-magnitude is important for the determination of the geometric albedo. We took the H-G solution from \citet{Oszkiewicz2011} with an absolute V-band magnitude of 5.85\,mag and the phase-slope parameter G of 0.12. The object's bolometric emissivity of 0.9 \citep{Landsman2018} was first translated into a hemispherical spectral emissivity of 0.9 at all wavelengths, and later on, into a wavelength-dependent solution.

\begin{figure}[!htb]
  \centering
\includegraphics[width=.50\textwidth]{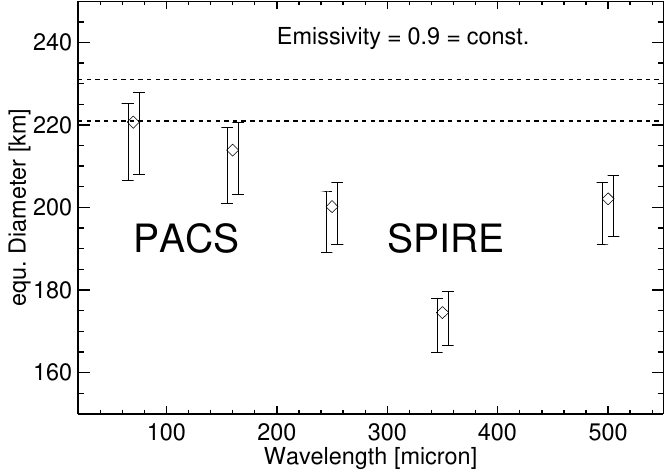}
\caption{Radiometric size solutions are shown for thermal inertias of 5\,J\,m$^{-2}$\,s$^{-1/2}$\,K$^{-1}$ (error bar left of reference wavelengths) and 150\,J\,m$^{-2}$\,s$^{-1/2}$\,K$^{-1}$ (error bar right of reference wavelengths). The error bars include the full range of ($\chi^2$-)acceptable solutions for all available measurements in that specific band, and considering low, intermediate, and high levels of surface roughness. The dashed line shows the published size ranges by \citet{Viikinkoski2018}, the diamond symbols correspond the low level of surface roughness \citep{Landsman2018} and a thermal inertia of 50\,J\,m$^{-2}$\,s$^{-1/2}$\,K$^{-1}$, which is intermediate between the ranges given in \citet{Matter2013} and \citet{Landsman2018}.}
\label{fig:radiometricDiameter}
\end{figure}

\begin{figure}[!htb]
  \centering
  \includegraphics[width=.50\textwidth]{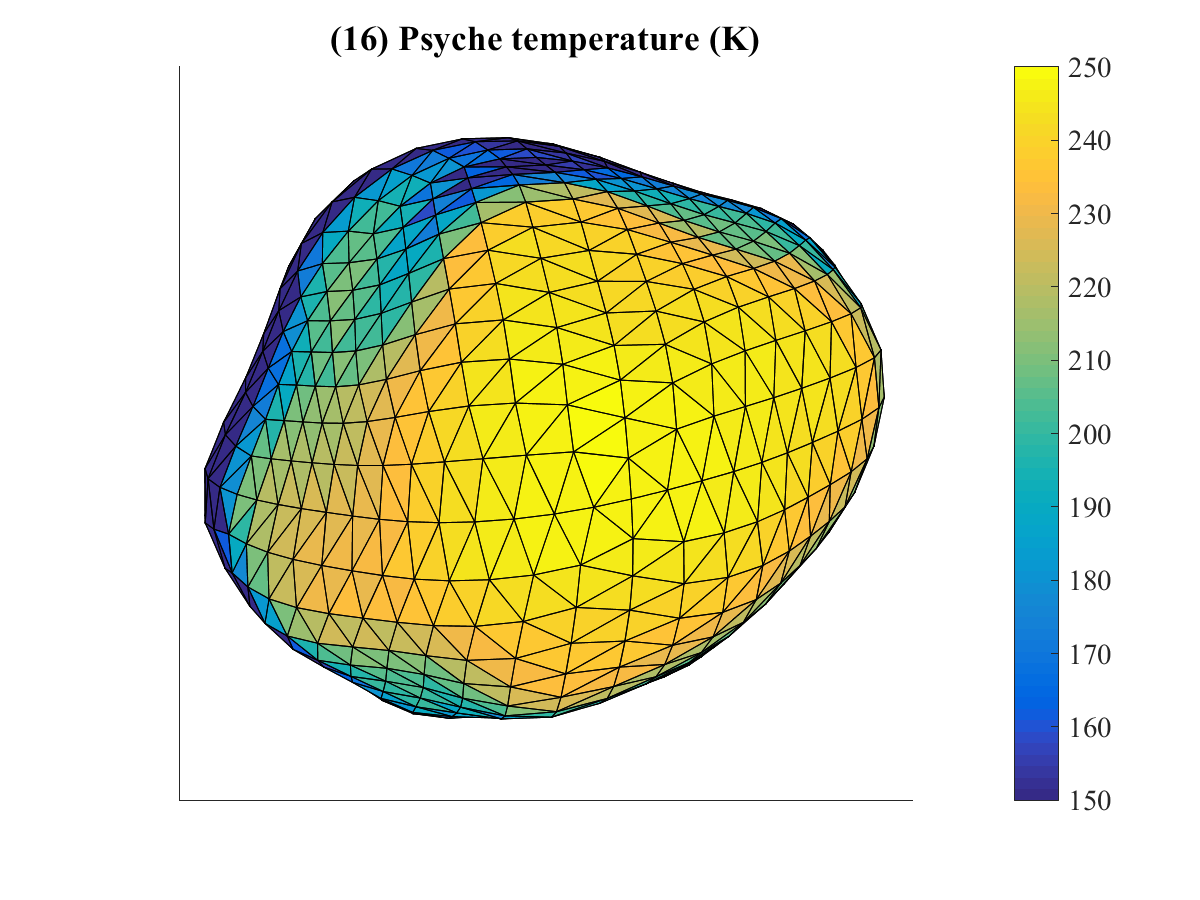}
  \caption{Psyche as seen from \herschel (in ecliptic-sky projection) during mid-time of the \herschel observing epochs. For the calculations, we used a thermal inertia of 50\,J\,m$^{-2}$\,s$^{-1/2}$\,K$^{-1}$ and a low level of surface roughness ($\rho$ = 0.1).}
  \label{fig:PsycheShapeTPM}
\end{figure}

We derived the radiometric size-albedo solutions for a wide range of $\Gamma$ and for low, intermediate, and high surface
roughness values ($\rho$ = 0.1, 0.4, 0.7) for all five PACS and SPIRE bands. The results for thermal inertias of 5 and 150\,J\,m$^{-2}$\,s$^{-1/2}$\,K$^{-1}$, each time for low, intermediate, and high levels of surface roughness, are shown in Fig.~\ref{fig:radiometricDiameter}. The figure shows: (i) the very small influence of both parameters on the radiometric size solution; (ii) a strong dependency of the derived radiometric size on the wavelengths; and (iii) the excellent agreement between the radiometric diameter at 70\,$\mu$m and the solution by \citet{Viikinkoski2018}.

We confirmed the strong wavelength-dependent diameter solution via a single CSO-SHARCII measurement (2007-Feb-12 11:21 UT, 350\,$\mu$m flux density 0.90 $\pm$ 0.09 Jy; D. Dowell, priv. comm. 2008). This led to a radiometric size in the range 167-199\,km, which is in fine agreement with the SPIRE 350\,$\mu$m result.

The small degree of influence on the part of surface roughness at these long wavelengths was expected from general radiometric studies \citep[Fig.~3 of][]{Muller2002}. For the thermal inertia, the reason is different: the specific Herschel viewing geometry towards the south pole region of Psyche means that there is effectively very little heat transport to the night side (the Sun illuminates the constantly visible South pole region, and only very small portions of other regions) and the thermal inertia is not well constrained by measurements with the given aspect angle. The wavelength-dependency of the radiometric diameter is a known effect \citep[see e.g.,][]{Muller1998,Muller2014} and it is caused by a change in the hemispherical spectral emissivity. We derived the Psyche-specific emissivity via calculated observation/TPM ratios.

We found a spectral emissivity of 0.9 out to about 100 \,$\mu$m, then a strong drop towards values of 0.6 at 350 \,$\mu$m and a subsequent increase to larger values towards 500 \,$\mu$m. However, this increase is only visible in the longest  wavelength SPIRE band, where the source is fainter than in the other bands.

In comparison with other large main-belt asteroids \citep{Muller1998,Muller2014}, Psyche shows a very strong and very unusual emissivity drop beyond 100\,$\mu$m, perhaps similar to what was found for
Vesta \citep{Muller1998}. If this emissivity drop is due to scattering processes by grains within the regolith, then our measurements would indicate large grain sizes of a few hundred micrometer in size.

For a better characterization of Psyche's peculiar emissivity behavior, more submm/mm measurements (as with ALMA bands 6-10) would be required. The \herschel measurements alone cannot constrain Psyche's thermal inertia or surface roughness due to the unfavorable close-to pole-on viewing geometry. In our second analysis, we combined our new measurements with the above-mentioned CSO data point and published thermal measurements from IRAS, AKARI, and WISE as provided by the asteroid thermal infrared database\footnote{\url{https://ird.konkoly.hu/}} \citep[][and references therein]{Szakats2020}. We also added the two rebinned Spitzer-IRS spectra (SL1 from 7.5-14\,$\mu$m from 2006 and SL1 from 5.2-8\,$\mu$m from 2004), as presented by \citet{Landsman2018}.

We repeated our radiometric analysis using \citet{Viikinkoski2018} spin-shape solution for the combined thermal IR data set, now applying the Vesta-like emissivity model \citep[see][Fig.4, lower panel]{Muller1998} to explain the PACS and SPIRE measurements.
We included all $\Gamma$ and roughness solutions where the radiometric size is within the published 224 $\pm$ 5\,km \citep{Viikinkoski2018} and for acceptable reduced $\chi^2$ values close to 1.0. We find that
thermal inertias between about 20 and 80\,J\,m$^{-2}$\,s$^{-1/2}$\,K$^{-1}$
and low levels of surface roughness $\rho < 0.4$ fit  the combined
thermal measurements best.

\begin{figure}[!htb]
  \centering
\includegraphics[width = 3in]{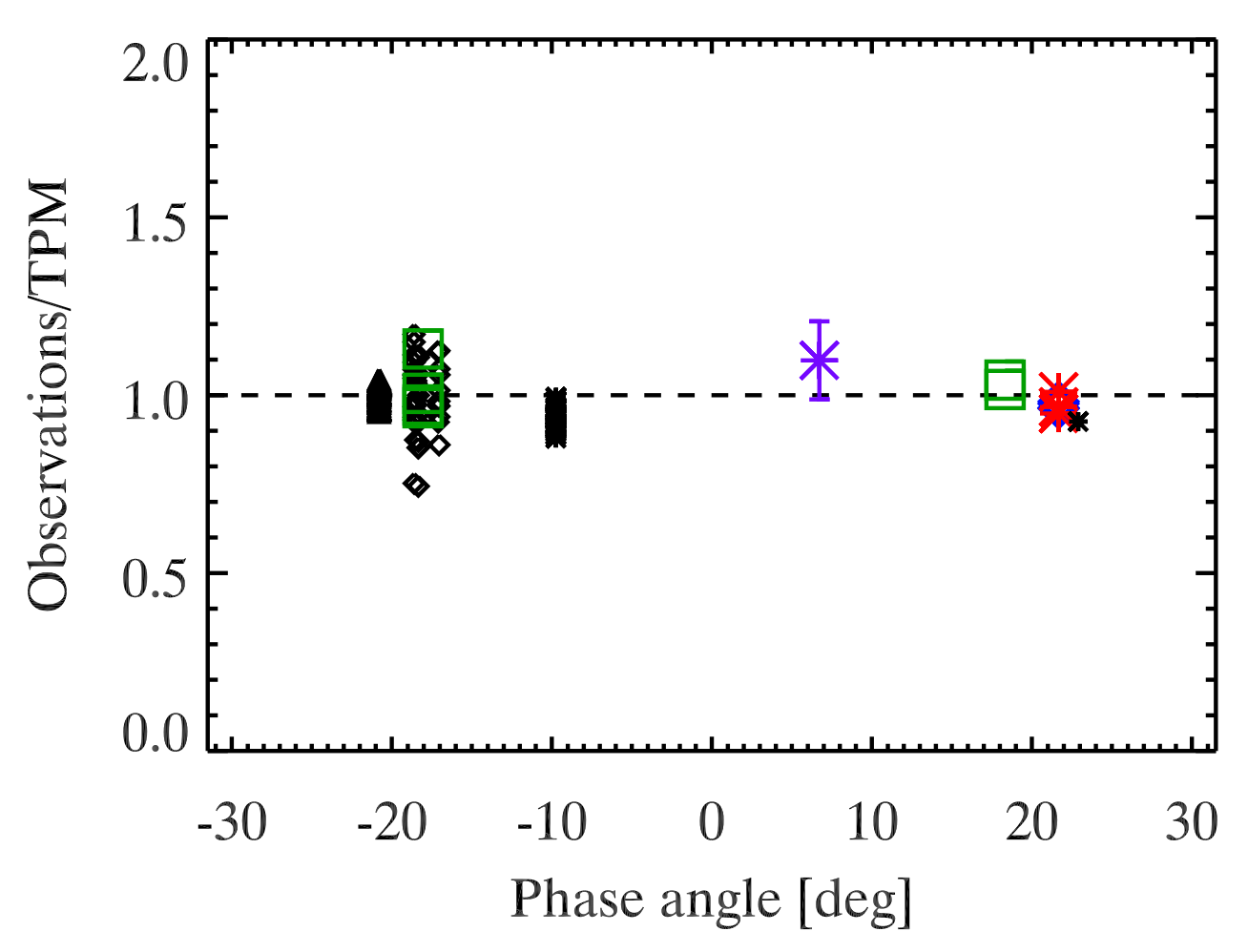}
\includegraphics[width = 3in]{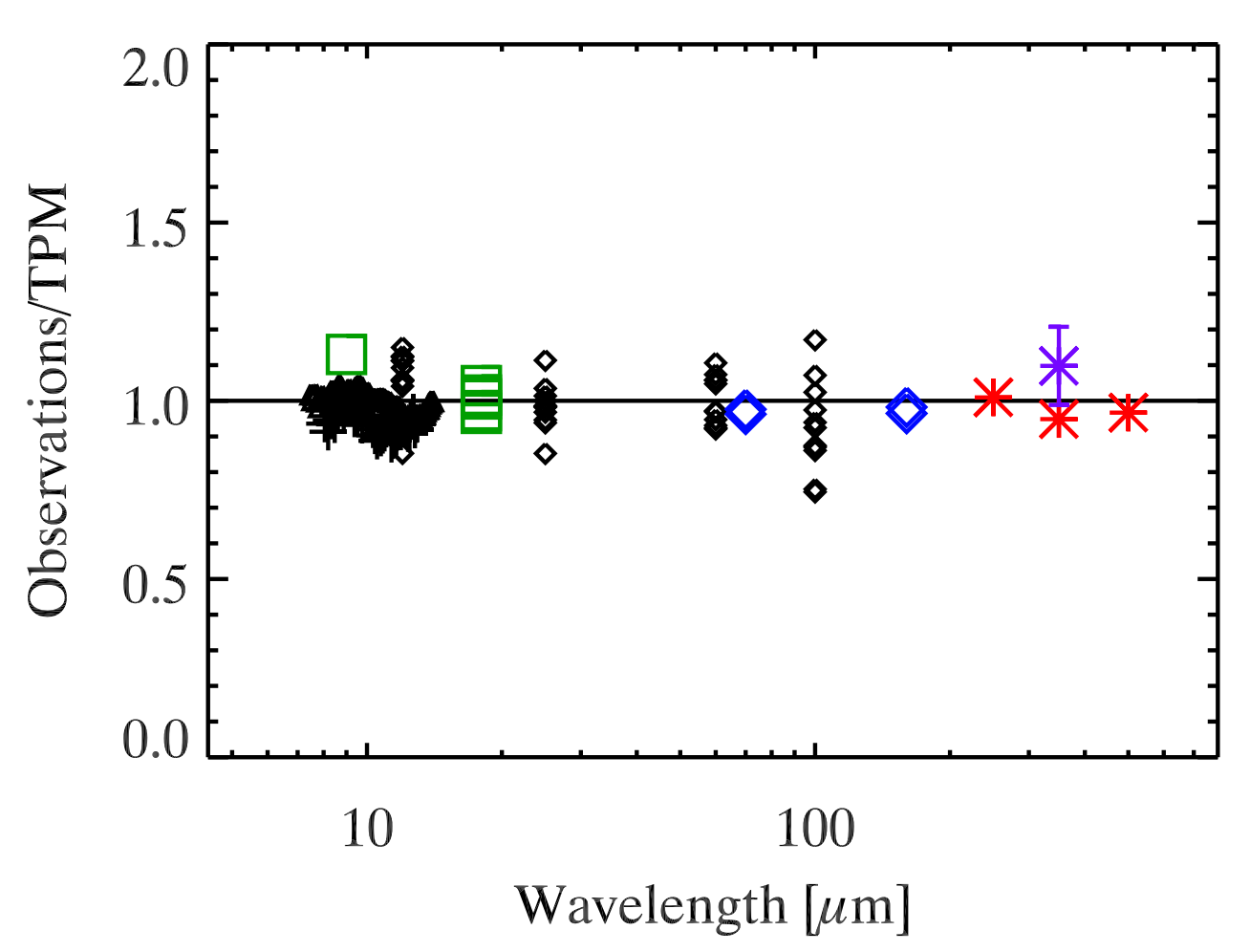}
 \caption{Thermal observations of Psyche divided by the corresponding thermophysical model predictions (using our best solution as given in the text). The new measurements are shown in color: green for the AKARI measurements, blue for the PACS, red for the SPIRE, and purple for the CSO data. The IRAS data are shown as diamonds, the Spitzer-IRS data as triangles, and the WISE data as plus symbols. The left figure shows the observation-to-model ratios as a function of phase angle (sensitive to thermal inertia) and the right figure shows it as function of wavelengths (sensitive to surface roughness and emissivity). Wrong settings in the model would show up as trends and slopes in these plots.}
 \label{fig:psyche_obsmod}
\end{figure}

Figure\ref{fig:psyche_obsmod} shows all available thermal measurements divided by the corresponding TPM predictions as a function of phase angle and wavelengths. For the TPM calculations, we used a size of 224 km, a thermal inertia of 50\,J\,m$^{-2}$\,s$^{-1/2}$\,K$^{-1}$, and a low level of surface roughness. The ratio plot as a function of phase angle is very sensitive to thermal inertia settings and lower or higher values would introduce a slope over this wide phase angle range. The ratio plot as a function of wavelength is indicative of surface roughness and emissivity properties. 

Our preference for low roughness is connected to the assumption of an emissivity of 0.9 at the shortest wavelengths below 10 \,$\mu$m. Higher values  for the surface roughness would be compatible with a lower emissivity at these short wavelengths below 10 \,$\mu$m (without violating our emissivity findings in the far-IR/submm range). This emissivity-roughness ambiguity is only relevant at wavelengths below the object's thermal emission peak, but cannot be solved with our limited data set.
 
The $\Gamma$ solution is in-between the findings by \citet{Matter2013} and by \citet{Landsman2018}.
A higher-quality solution for Psyche's thermal properties would require more thermal measurements in an equatorial view and preferentially for a wide range of phase angles. Unfortunately, the \citet{Matter2013} data are not available in a tabulated form. These fluxes have been calibrated under the assumptions of a different spin-shape solution and their usability for standard radiometric studies is not clear. However, we have made predictions with our best radiometric solution for these observing epochs and we found a 10-20\% systematic offset with respect to the VLT-MIDI measurements, which is not compatible with the rest of the thermal measurements.

%----------------------------------------------------------------------------------------------------%
\section{Conclusions}\label{sec:conclusions}

In this work, we present the ESASky Solar System Object Search Service, developed in the ESAC Science Data Centre and included in the ESASky application, with the intent of facilitating the scientific exploitation of the ESA archival astronomy data holdings by the Solar System community. A set of all the positional cross-matches and uncertainties, apparent magnitudes and distances, of all asteroids included in the \astorb database by July 2019 with respect to all the high-level imaging observations stored at the ESDC Astronomy Archives for the Hubble Space Telescope, \herschel, and XMM-Newton Observatory is provided. Together with the results currently provided by this service, we included three catalogs with the selected potential detections with magnitudes or thermal fluxes above instrumental sensitivity.

The caveats introduced in this work were described in Section \ref{sec:results}, linked to the nature of the stored metadata, in particular, to the time-related information available for high-level processed data and its peculiarities across different missions.
Finally, to showcase the potential of this service, we included a thermophysical analysis for the mission-target asteroid (16) Psyche.

Our \herschel PACS and SPIRE serendipitous detections helped to settle a long-standing discussion on the object's thermal inertia, which was found to be between 20 and 80\,J\,m$^{-2}$\,s$^{-1/2}$\,K$^{-1}$, intermediate between the previous conflicting determinations and more consistent with silicate powder rather than with a metallic surface. We can also put constraints on the surface roughness properties (rms of surface slopes below 0.4) and the hemispherical emissivity in the far-IR and submm range, which is found to be very low at 350\,$\mu$m. The far-infrared emissivity curve, similar to that of Vesta, may also favor a high silicate content. Similar radiometric studies can be expected for several other small bodies serendipitously detected by \herschel.
   
Future releases of the SSOSS will include this computation of the apparent magnitude of the SSOs at the wavelength of the observation based on the work presented here. A new interface will be developed to allow user-defined orbital parameters as input and on-the-fly computation of potential detections per object and mission, in addition to more missions being included.

%-------------------------------------------------------------------
\begin{acknowledgements}
MM acknowledges funding by the European Space Agency under the research contract C4000122918.
TM has received funding from the European Union\'s Horizon 2020 Research and Innovation Programme, under Grant Agreement no 687378, as part of the project "Small Bodies Near and Far" (SBNAF).BC and JB acknowledge support from the ESAC Faculty Visitor Programme. This research has made use of the SVO Filter Profile Service\footnote{\url{http://svo2.cab.inta-csic.es/theory/fps/}} supported from the Spanish MINECO through grant AYA2017-84089.
\end{acknowledgements}
%-------------------------------------------------------------------
\bibliographystyle{./bibtex/aa}
\interlinepenalty=10000
\bibliography{./bibtex/biblio.bib}

%-------------------------------------------------------------------
\begin{appendix}
\section{Geometrical cross-match types}\label{cross-match descriptions}

Given \(sso_{start}\) and \(sso_{stop}\) as the sky positions computed by the Eproc software for a given $t_{start}$ and $t_{stop}$ of a given observation, the pipeline performs a series of geometrical tests to verify whether there is a cross-match between the observation FoV polygon in sky coordinates and the \(SSO_{start}\) and \(SSO_{stop}\) positions with their uncertainties. These positions are represented with geometrical circles centered at each position with a radius equal to the position uncertainty.

Three types of geometrical cross-matches have been identified, along with four different algorithms: two for the cross-match type 1, one for the type 2, and one for the type 3. The pipeline runs the cross-match checks in a specific order, first computing  the algorithms with less complexity and computational time cost, from type 2 -> type 1 -> type 3.

In the code, there are three main blocks described in the pseudo-code below. As soon as one cross-match test succeeds, the pipeline exits with a positive value and the cross-match result is saved in the database, including the object details, associated observation metadata, and cross-match type.

\subsection{Cross-match type 1}
This category includes all cross-matches where at least one of the geometrical circles representing the SSO (start or stop positions) intersects with one of the polygon segments representing the observation FoV. This family of cross-matches is identified by two separate algorithms or sub-types in the pipeline, so-called types 1.1 and 1.2.

\begin{figure}[!htb]
  \centering
  \includegraphics[width=.24\textwidth]{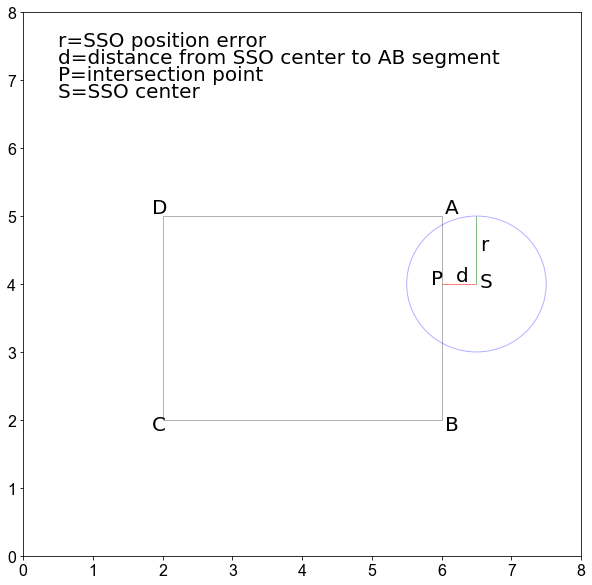}
  \includegraphics[width=.24\textwidth]{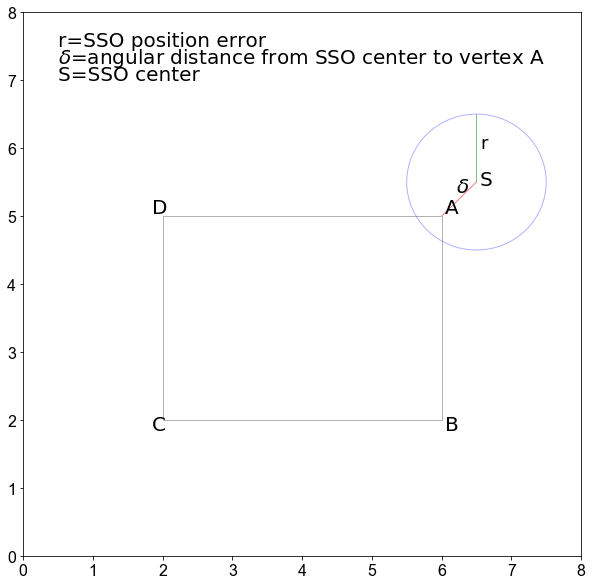}
  \caption{Cross-match subtypes 1.1, (left panel) and 1.2 (right panel).}
  \label{fig:xmatch_t1_2}
\end{figure}

Cross-match type 1.1 tests whether the distance from the object's centre to a given polygon segment is less than the computed position uncertainty, that is, whether the intersection point P between the perpendicular to the FoV, starting from the SSO centre, belongs to the FoV. Whereas when the angular distance between the position of the object and one of the two vertices of the polygon is less than the position uncertainty, the cross-match is of type 1.2. (Fig.~\ref{fig:xmatch_t1_2}). Both types 1.1 and 1.2 have been marked as cross-match type 1 in the results provided by the pipeline.

\subsection{Cross-match type 2}
\begin{figure}[!htb]
  \centering
  \includegraphics[width=.24\textwidth]{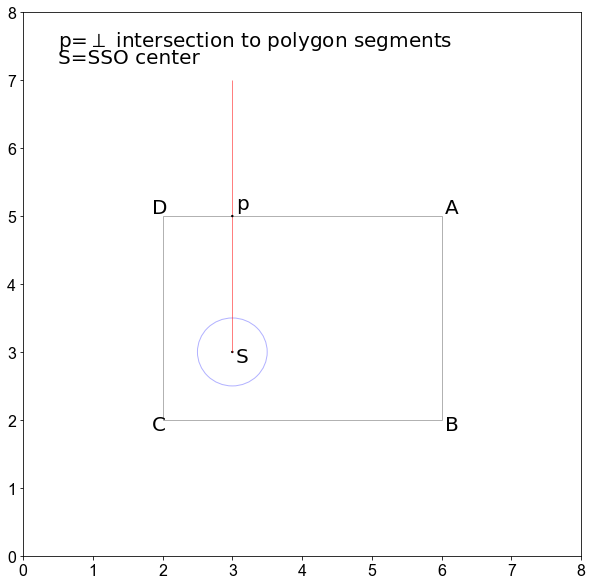}
\includegraphics[width=.24\textwidth]{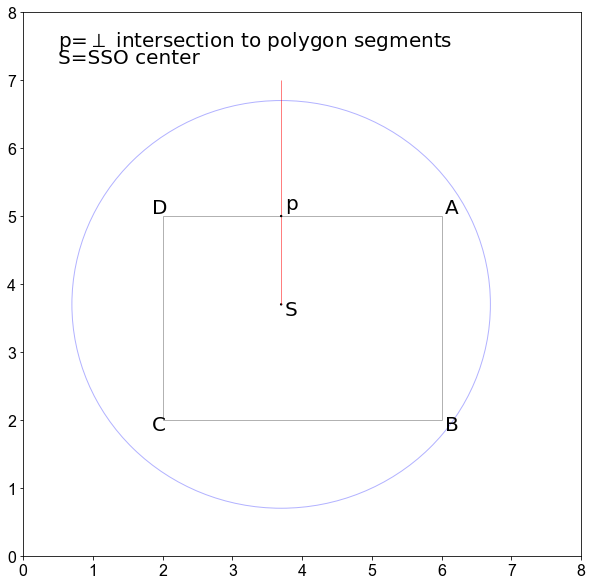}
  \caption{Cross-match type 2, where the SSO position lies inside a given FoV, regardless of the SSO uncertainty radius.}
  \label{fig:xmatch_t2b}
\end{figure}

This is the case when one of the SSO centers lies inside the footprint (Fig.~\ref{fig:xmatch_t2b}). The geometrical check is performed by counting the number of the intersections between a line segment originating from the centre of the SSO and the FoV polygon segments. Odd intersections means positive cross-match of type 2.

\subsection{Cross-match type 3}
\begin{figure}[!htb]
  \centering
  \includegraphics[width=.25\textwidth]{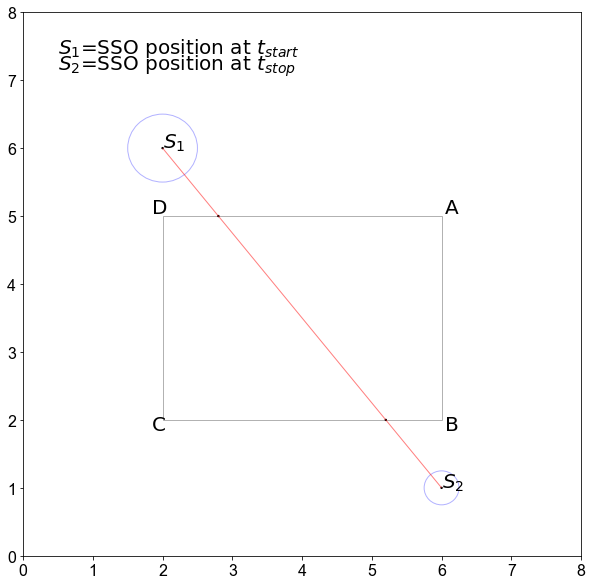}
  \smallskip
  \caption{Cross-match type 3.}
  \label{fig:xmatch_t3}
\end{figure}

Cross-match type 3 is the last step in the chain of geometrical algorithms, where none of the SSO calculated positions and uncertainties overlap with the observation FoV. It calculates whether there is an intersection between the line crossing both $SSO_{start}$ and $SSO_{stop}$ positions with any of the FoV polygon segments (Fig.~\ref{fig:xmatch_t3}).

\section{List of potential detections}

The full store of tables with the list of theoretical detections used in this work is available through the ESASky TAP as described in Section \ref{sec:results}, via the sso\_20190705 schema, and table names xmatch\_<mission>. Here, we introduce a subsample of the first detection results for each mission for illustrative purposes.

%----------------------------------------------------------------------------------------------------%
\longtab[1]{    
\tiny
\begin{longtable}{p{2.2cm}ccccccccc}
\caption{{\it Herschel Serendipitous Detection List of asteroids}. A sample of the brightest 100 serendipitous detections with a theoretical thermal flux at 70$\mu$m above \textit{Herschel} sensitivity and cross-match type 2 (see Appendix \ref{cross-match descriptions}).}
\label{table:herschelXmatch_list}\\
\hline
\hline
Asteroid Id & Observation Id &  RA$_1$ & Dec$_1$\tablefootmark{a} & RA$_2$ & Dec$_2$\tablefootmark{b} & $\delta$Pos\tablefootmark{c} &  m$_{v}$ &F$_{70}$\tablefootmark{d}&d\tablefootmark{e} \\
 & & \multicolumn{2}{c}{\it (J2000.0)}&\multicolumn{2}{c}{\it (J2000.0)} & \it{(arcsec)} & & \it{(Jy)}&\it{(AU)}\\
\endfirsthead
\caption{Continued.} \\
\hline
Asteroid Id & Observation Id &  RA$_1$ & Dec$_1$\tablefootmark{a} & RA$_2$ & Dec$_2$\tablefootmark{b} & $\delta$Pos\tablefootmark{c}&  m$_{v}$ &F$_{70}$\tablefootmark{d} & d\tablefootmark{e} \\
 & & \multicolumn{2}{c}{\it (J2000.0)}&\multicolumn{2}{c}{\it (J2000.0)} & \it{(arcsec)} & & \it{(Jy)} & \it{(AU)}\\
\hline
\endhead
\hline
\endfoot
\hline
\endlastfoot
\hline
(16) Psyche&1342202251&04:27:08.97&18$^{\circ}$57'01.68"&04:27:08.97&18$^{\circ}$57'01.68"&0.58&11.14&48.940&02.78\\
(16) Psyche&1342202250&04:27:03.80&18$^{\circ}$56'52.81"&04:27:03.80&18$^{\circ}$56'52.81"&0.58&11.14&48.920&02.78\\
(13) Egeria&1342204100&17:12:17.60&-39$^{\circ}$02'45.98"&17:12:17.60&-39$^{\circ}$02'45.98"&0.41&11.97&41.870&02.48\\
(13) Egeria&1342204101&17:12:24.05&-39$^{\circ}$02'29.61"&17:12:24.05&-39$^{\circ}$02'29.61"&0.41&11.98&41.810&02.49\\
(107) Camilla&1342251927&06:24:05.98&12$^{\circ}$42'25.63"&06:24:05.98&12$^{\circ}$42'25.63"&0.03&13.15&27.830&03.11\\
(107) Camilla&1342251926&06:24:01.81&12$^{\circ}$42'49.18"&06:24:01.81&12$^{\circ}$42'49.18"&0.03&13.15&27.800&03.11\\
(87) Sylvia&1342250801&05:56:29.40&23$^{\circ}$21'26.14"&05:56:29.40&23$^{\circ}$21'26.14"&0.02&13.29&25.000&03.54\\
(87) Sylvia&1342250800&05:56:23.60&23$^{\circ}$21'14.20"&05:56:23.60&23$^{\circ}$21'14.20"&0.02&13.29&24.970&03.54\\
(393) Lampetia&1342185641&18:34:27.97&-07$^{\circ}$32'25.46"&18:34:27.97&-07$^{\circ}$32'25.46"&0.14&12.31&18.010&01.75\\
(393) Lampetia&1342185642&18:34:44.33&-07$^{\circ}$32'58.29"&18:34:44.33&-07$^{\circ}$32'58.29"&0.14&12.31&17.990&01.75\\
(121) Hermione&1342190654&04:34:21.71&24$^{\circ}$12'21.64"&04:34:21.71&24$^{\circ}$12'21.64"&0.19&13.07&15.490&02.85\\
(675) Ludmilla&1342202254&04:27:57.22&27$^{\circ}$27'08.45"&04:27:57.22&27$^{\circ}$27'08.45"&0.33&12.72&15.480&02.44\\
(121) Hermione&1342190655&04:34:23.68&24$^{\circ}$12'30.56"&04:34:23.68&24$^{\circ}$12'30.56"&0.19&13.07&15.470&02.85\\
(690) Wratislavia&1342202254&04:24:36.50&26$^{\circ}$10'30.25"&04:24:36.50&26$^{\circ}$10'30.25"&0.28&13.47&13.200&02.90\\
(84) Klio&1342204366&17:46:22.24&-29$^{\circ}$07'18.84"&17:46:22.24&-29$^{\circ}$07'18.84"&0.07&12.92&10.530&01.48\\
(212) Medea&1342202253&04:31:29.88&25$^{\circ}$48'02.57"&04:31:29.88&25$^{\circ}$48'02.57"&0.12&13.91&08.616&03.06\\
(212) Medea&1342202252&04:31:19.85&25$^{\circ}$47'35.66"&04:31:19.85&25$^{\circ}$47'35.66"&0.12&13.91&08.609&03.06\\
(43) Ariadne&1342216014&17:31:03.27&-25$^{\circ}$24'56.89"&17:31:03.27&-25$^{\circ}$24'56.89"&0.04&11.85&07.808&01.65\\
(43) Ariadne&1342216013&17:30:50.27&-25$^{\circ}$24'50.86"&17:30:50.27&-25$^{\circ}$24'50.86"&0.04&11.85&07.795&01.65\\
(554) Peraga&1342202253&04:42:30.81&24$^{\circ}$44'34.19"&04:42:30.81&24$^{\circ}$44'34.19"&0.38&13.72&07.017&02.28\\
(241) Germania&1342214762&17:54:37.76&-25$^{\circ}$04'10.23"&17:54:37.76&-25$^{\circ}$04'10.23"&0.04&13.53&06.521&03.32\\
(241) Germania&1342214761&17:54:29.28&-25$^{\circ}$04'12.65"&17:54:29.28&-25$^{\circ}$04'12.65"&0.04&13.53&06.515&03.32\\
(388) Charybdis&1342204366&17:47:03.21&-30$^{\circ}$03'15.52"&17:47:03.21&-30$^{\circ}$03'15.52"&0.14&14.01&05.386&02.54\\
(388) Charybdis&1342204367&17:47:08.81&-30$^{\circ}$02'57.96"&17:47:08.81&-30$^{\circ}$02'57.96"&0.14&14.01&05.379&02.54\\
(58) Concordia&1342185647&06:11:39.03&17$^{\circ}$19'13.87"&06:11:39.03&17$^{\circ}$19'13.87"&0.08&13.88&04.970&02.29\\
(58) Concordia&1342185646&06:11:36.29&17$^{\circ}$19'27.69"&06:11:36.29&17$^{\circ}$19'27.69"&0.08&13.88&04.964&02.29\\
(790) Pretoria&1342202254&04:20:47.91&28$^{\circ}$06'31.61"&04:20:47.91&28$^{\circ}$06'31.61"&0.13&14.57&04.590&03.88\\
(790) Pretoria&1342202090&04:16:17.19&28$^{\circ}$01'15.06"&04:16:17.19&28$^{\circ}$01'15.06"&0.13&14.58&04.470&03.93\\
(683) Lanzia&1342263847&17:19:17.61&-28$^{\circ}$13'01.25"&17:19:17.61&-28$^{\circ}$13'01.25"&0.34&14.59&02.240&03.24\\
(683) Lanzia&1342263846&17:19:09.27&-28$^{\circ}$13'06.39"&17:19:09.27&-28$^{\circ}$13'06.39"&0.34&14.59&02.239&03.24\\
(674) Rachele&1342214578&16:39:53.27&-21$^{\circ}$02'12.88"&16:39:53.27&-21$^{\circ}$02'12.88"&0.05&13.46&02.168&03.25\\
(674) Rachele&1342214577&16:39:44.40&-21$^{\circ}$01'34.84"&16:39:44.40&-21$^{\circ}$01'34.84"&0.05&13.46&02.165&03.26\\
(199) Byblis&1342185643&17:44:15.75&-27$^{\circ}$54'47.35"&17:44:15.75&-27$^{\circ}$54'47.35"&0.26&13.92&01.885&02.87\\
(628) Christine&1342190614&04:23:37.69&15$^{\circ}$57'16.43"&04:23:37.69&15$^{\circ}$57'16.43"&0.12&14.16&01.853&02.24\\
(366) Vincentina&1342214714&17:35:03.26&-33$^{\circ}$05'11.54"&17:35:03.26&-33$^{\circ}$05'11.54"&0.04&14.62&01.711&03.25\\
(142) Polana&1342190652&04:40:00.07&23$^{\circ}$05'10.06"&04:40:00.07&23$^{\circ}$05'10.06"&0.69&15.02&01.666&02.12\\
(142) Polana&1342190653&04:40:02.09&23$^{\circ}$05'09.72"&04:40:02.09&23$^{\circ}$05'09.72"&0.69&15.02&01.665&02.12\\
(634) Ute&1342188084&23:03:50.42&-16$^{\circ}$30'32.56"&23:03:50.42&-16$^{\circ}$30'32.56"&0.13&14.73&01.614&02.47\\
(634) Ute&1342188085&23:04:00.19&-16$^{\circ}$29'26.71"&23:04:00.19&-16$^{\circ}$29'26.71"&0.13&14.73&01.612&02.47\\
(634) Ute&1342188086&23:04:09.96&-16$^{\circ}$28'20.77"&23:04:09.96&-16$^{\circ}$28'20.77"&0.13&14.73&01.610&02.47\\
(634) Ute&1342188087&23:04:19.75&-16$^{\circ}$27'14.74"&23:04:19.75&-16$^{\circ}$27'14.74"&0.13&14.74&01.608&02.47\\
(1212) Francette&1342267755&16:49:29.25&-14$^{\circ}$00'39.30"&16:49:29.25&-14$^{\circ}$00'39.30"&0.06&16.10&01.487&03.58\\
(1212) Francette&1342267754&16:49:27.70&-14$^{\circ}$00'48.29"&16:49:27.70&-14$^{\circ}$00'48.29"&0.06&16.10&01.486&03.58\\
(509) Iolanda&1342218645&18:29:40.60&-11$^{\circ}$07'23.48"&18:29:40.60&-11$^{\circ}$07'23.48"&0.02&13.99&01.187&02.71\\
(509) Iolanda&1342218644&18:29:38.31&-11$^{\circ}$08'04.83"&18:29:38.31&-11$^{\circ}$08'04.83"&0.02&13.99&01.186&02.71\\
(509) Iolanda&1342218642&18:29:33.52&-11$^{\circ}$09'30.64"&18:29:33.52&-11$^{\circ}$09'30.64"&0.02&13.99&01.183&02.71\\
(721) Tabora&1342204366&17:42:41.12&-31$^{\circ}$25'20.91"&17:42:41.12&-31$^{\circ}$25'20.91"&0.17&15.60&00.972&03.39\\
(721) Tabora&1342204367&17:42:44.72&-31$^{\circ}$25'11.74"&17:42:44.72&-31$^{\circ}$25'11.74"&0.17&15.60&00.971&03.40\\
(721) Tabora&1342204368&17:42:48.08&-31$^{\circ}$25'03.22"&17:42:48.08&-31$^{\circ}$25'03.22"&0.17&15.60&00.970&03.40\\
(338) Budrosa&1342204088&16:39:16.33&-23$^{\circ}$51'51.82"&16:39:16.33&-23$^{\circ}$51'51.82"&0.23&14.09&00.934&02.80\\
(338) Budrosa&1342204089&16:39:18.79&-23$^{\circ}$51'51.76"&16:39:18.79&-23$^{\circ}$51'51.76"&0.23&14.09&00.933&02.80\\
(816) Juliana&1342218643&18:33:03.77&-10$^{\circ}$34'47.75"&18:33:03.77&-10$^{\circ}$34'47.75"&0.04&15.99&00.908&02.63\\
(816) Juliana&1342218642&18:33:01.32&-10$^{\circ}$35'01.87"&18:33:01.32&-10$^{\circ}$35'01.87"&0.04&15.99&00.907&02.63\\
(830) Petropolitana&1342190616&04:27:44.04&26$^{\circ}$05'33.30"&04:27:44.04&26$^{\circ}$05'33.30"&0.55&14.53&00.787&02.60\\
(475) Ocllo&1342204858&04:43:55.11&25$^{\circ}$25'40.47"&04:43:55.11&25$^{\circ}$25'40.47"&0.04&15.23&00.750&01.35\\
(1237) Genevieve&1342204368&17:39:43.44&-32$^{\circ}$01'16.86"&17:39:43.44&-32$^{\circ}$01'16.86"&0.44&15.67&00.702&02.23\\
(1237) Genevieve&1342204369&17:39:50.33&-32$^{\circ}$01'11.63"&17:39:50.33&-32$^{\circ}$01'11.63"&0.44&15.68&00.701&02.23\\
(1017) Jacqueline&1342267726&16:48:42.97&-13$^{\circ}$14'14.32"&16:48:42.97&-13$^{\circ}$14'14.32"&0.09&15.67&00.686&02.01\\
(659) Nestor&1342202253&04:40:33.51&26$^{\circ}$12'39.24"&04:40:33.51&26$^{\circ}$12'39.24"&0.28&16.89&00.582&05.49\\
(659) Nestor&1342202252&04:40:28.66&26$^{\circ}$12'24.74"&04:40:28.66&26$^{\circ}$12'24.74"&0.28&16.89&00.582&05.49\\
(118) Peitho&1342204368&17:36:55.79&-30$^{\circ}$41'06.56"&17:36:55.79&-30$^{\circ}$41'06.56"&0.10&14.39&00.541&02.50\\
(118) Peitho&1342204369&17:37:01.25&-30$^{\circ}$40'58.99"&17:37:01.25&-30$^{\circ}$40'58.99"&0.10&14.39&00.541&02.50\\
(1116) Catriona&1342250343&05:22:04.80&37$^{\circ}$03'45.29"&05:22:04.80&37$^{\circ}$03'45.29"&0.26&14.61&00.523&02.36\\
(1116) Catriona&1342250342&05:21:50.46&37$^{\circ}$02'47.71"&05:21:50.46&37$^{\circ}$02'47.71"&0.26&14.62&00.523&02.36\\
(1116) Catriona&1342250233&05:19:11.00&36$^{\circ}$52'05.15"&05:19:11.00&36$^{\circ}$52'05.15"&0.26&14.63&00.516&02.38\\
(1116) Catriona&1342250232&05:18:56.63&36$^{\circ}$51'07.05"&05:18:56.63&36$^{\circ}$51'07.05"&0.26&14.63&00.515&02.38\\
(1471) Tornio&1342214504&03:41:17.92&32$^{\circ}$37'23.99"&03:41:17.92&32$^{\circ}$37'23.99"&0.23&15.79&00.507&02.22\\
(1471) Tornio&1342214505&03:41:29.16&32$^{\circ}$37'02.11"&03:41:29.16&32$^{\circ}$37'02.11"&0.23&15.79&00.507&02.22\\
(1524) Joensuu&1342239278&04:23:52.26&36$^{\circ}$16'11.18"&04:23:52.26&36$^{\circ}$16'11.18"&0.16&16.25&00.485&02.69\\
(1524) Joensuu&1342239279&04:23:58.85&36$^{\circ}$15'23.91"&04:23:58.85&36$^{\circ}$15'23.91"&0.16&16.25&00.484&02.70\\
(436) Patricia&1342213183&15:38:36.55&-33$^{\circ}$28'45.57"&15:38:36.55&-33$^{\circ}$28'45.57"&0.18&16.35&00.449&03.73\\
(436) Patricia&1342213182&15:38:29.24&-33$^{\circ}$27'54.36"&15:38:29.24&-33$^{\circ}$27'54.36"&0.18&16.35&00.448&03.73\\
(1254) Erfordia&1342205093&16:23:37.42&-24$^{\circ}$08'37.80"&16:23:37.42&-24$^{\circ}$08'37.80"&0.02&16.55&00.448&03.41\\
(1254) Erfordia&1342205094&16:23:52.22&-24$^{\circ}$08'51.63"&16:23:52.22&-24$^{\circ}$08'51.63"&0.02&16.55&00.447&03.41\\
(969) Leocadia&1342204859&04:50:30.64&25$^{\circ}$15'39.12"&04:50:30.64&25$^{\circ}$15'39.12"&0.12&16.26&00.446&01.51\\
(969) Leocadia&1342204858&04:50:26.85&25$^{\circ}$15'31.56"&04:50:26.85&25$^{\circ}$15'31.56"&0.12&16.26&00.445&01.51\\
(1280) Baillauda&1342202254&04:25:31.50&26$^{\circ}$58'02.81"&04:25:31.50&26$^{\circ}$58'02.81"&0.22&16.51&00.436&03.52\\
(1771) Makover&1342184474&17:47:06.85&-26$^{\circ}$59'08.95"&17:47:06.85&-26$^{\circ}$59'08.95"&0.25&16.49&00.426&03.28\\
(352) Gisela&1342184488&07:29:39.01&21$^{\circ}$04'41.10"&07:29:39.01&21$^{\circ}$04'41.10"&0.04&14.22&00.395&01.96\\
(1392) Pierre&1342202090&04:14:48.02&28$^{\circ}$50'46.28"&04:14:48.02&28$^{\circ}$50'46.28"&0.04&16.33&00.385&02.32\\
(214) Aschera&1342190617&04:09:53.35&25$^{\circ}$03'46.87"&04:09:53.35&25$^{\circ}$03'46.87"&0.12&14.06&00.347&02.15\\
(214) Aschera&1342190618&04:09:56.91&25$^{\circ}$03'48.07"&04:09:56.91&25$^{\circ}$03'48.07"&0.12&14.07&00.346&02.15\\
(1166) Sakuntala&1342218647&18:32:33.75&-08$^{\circ}$03'59.98"&18:32:33.75&-08$^{\circ}$03'59.98"&0.40&14.29&00.341&01.61\\
(1166) Sakuntala&1342218646&18:32:27.55&-08$^{\circ}$03'59.28"&18:32:27.55&-08$^{\circ}$03'59.28"&0.40&14.29&00.341&01.61\\
(13832)1999 XR13&1342183070&17:45:05.35&-29$^{\circ}$15'12.37"&17:45:05.35&-29$^{\circ}$15'12.37"&0.04&16.43&00.332&02.82\\
(13832)1999 XR13&1342183071&17:45:06.63&-29$^{\circ}$15'17.07"&17:45:06.63&-29$^{\circ}$15'17.07"&0.04&16.43&00.332&02.82\\
(390) Alma&1342190327&03:30:10.80&29$^{\circ}$07'26.53"&03:30:10.80&29$^{\circ}$07'26.53"&0.03&15.04&00.322&02.09\\
(3815) Konig&1342183068&18:27:43.23&-11$^{\circ}$58'56.03"&18:27:43.23&-11$^{\circ}$58'56.03"&0.13&16.89&00.319&01.93\\
(3815) Konig&1342183069&18:27:43.97&-11$^{\circ}$59'05.08"&18:27:43.97&-11$^{\circ}$59'05.08"&0.13&16.89&00.319&01.93\\
(394) Arduina&1342190654&04:27:22.69&24$^{\circ}$28'10.89"&04:27:22.69&24$^{\circ}$28'10.89"&0.01&14.87&00.318&02.44\\
(394) Arduina&1342190655&04:27:25.47&24$^{\circ}$28'19.87"&04:27:25.47&24$^{\circ}$28'19.87"&0.01&14.87&00.318&02.44\\
(908) Buda&1342204102&17:40:13.97&-28$^{\circ}$51'09.75"&17:40:13.97&-28$^{\circ}$51'09.75"&0.26&15.92&00.288&02.46\\
(908) Buda&1342204103&17:40:17.79&-28$^{\circ}$51'18.42"&17:40:17.79&-28$^{\circ}$51'18.42"&0.26&15.93&00.287&02.46\\
(1687) Glarona&1342184472&19:08:41.24&-23$^{\circ}$57'39.38"&19:08:41.24&-23$^{\circ}$57'39.38"&0.36&16.01&00.280&02.87\\
(1687) Glarona&1342184473&19:08:44.78&-23$^{\circ}$57'34.48"&19:08:44.78&-23$^{\circ}$57'34.48"&0.36&16.01&00.279&02.87\\
(292) Ludovica&1342214578&16:31:36.73&-22$^{\circ}$43'48.48"&16:31:36.73&-22$^{\circ}$43'48.48"&0.53&14.99&00.277&02.48\\
(292) Ludovica&1342214577&16:31:23.62&-22$^{\circ}$42'35.15"&16:31:23.62&-22$^{\circ}$42'35.15"&0.53&14.99&00.277&02.48\\
(908) Buda&1342204366&17:43:23.10&-28$^{\circ}$57'30.79"&17:43:23.10&-28$^{\circ}$57'30.79"&0.26&15.99&00.272&02.53\\
(908) Buda&1342204367&17:43:28.03&-28$^{\circ}$57'39.63"&17:43:28.03&-28$^{\circ}$57'39.63"&0.26&15.99&00.272&02.53\\
(2967) Vladisvyat&1342250333&05:38:54.92&32$^{\circ}$46'03.03"&05:38:54.92&32$^{\circ}$46'03.03"&0.08&16.75&00.242&03.03\\
\end{longtable}
\tablefoot{SPIRE/PACS Parallel Observing Mode in all observations listed.\\
\tablefoottext{a}{Predicted Right Ascension (RA, J2000) and Declination (Dec, J2000) at the start of the observation}
\tablefoottext{b}{Predicted Right Ascension (RA, J2000) and Declination (Dec, J2000) at the end of the exposure time}
\tablefoottext{c}{Propagated position uncertainty}
\tablefoottext{d}{Theoretical thermal Flux computed at 70$\mu$m.}
\tablefoottext{e}{Distance from the satellite at the time of the observation}}
}
\bigskip % just to make a space here
%----------------------------------------------------------------------------------------------------%
%----------------------------------------------------------------------------------------------------%
\longtab[2]{    
\begin{longtable}{p{2.4cm}clccccccc}
\caption{{\it HST Serendipitous Detections of Asteroids}. A sample of the brightest 100 serendipitous detections of asteroids from HST, selected with $\delta$Pos < 5 arcsec, and cross-match type 2.}
\label{table:hstlXmatch_list}\\
\hline
\hline
Asteroid Id & Observation Id & Instrument Mode &  RA$_1$ & Dec$_1$\tablefootmark{a} & RA$_2$ & Dec$_2$\tablefootmark{b} & $\delta$Pos\tablefootmark{c}&  m$_{v}$ & d\tablefootmark{d} \\
 & & & \multicolumn{2}{c}{\it (J2000.0)}&\multicolumn{2}{c}{\it (J2000.0)} & \it{(arcsec)} & & \it{(AU)}\\
\hline
\\
\endfirsthead
\caption{Continued.} \\
\hline
Asteroid Id & Observation Id & Instrument Mode &  RA$_1$ & Dec$_1$\tablefootmark{a} & RA$_2$ & Dec$_2$\tablefootmark{b} & $\delta$Pos\tablefootmark{c}&  m$_{v}$ & d\tablefootmark{d}\\
 & & & \multicolumn{2}{c}{\it (J2000.0)}&\multicolumn{2}{c}{\it (J2000.0)} & \it{(arcsec)} & & \it{(AU)}\\
\hline
\\
\endhead
\hline
\endfoot
\hline
\\
\endlastfoot
(507) Laodica&j8pu10010&ACS/WFC&10:01:33.35&02$^{\circ}$19'18.19"&10:01:33.35&02$^{\circ}$19'18.19"&0.60&14.12&02.36\\
(490) Veritas&j9qc06010&ACS/WFC&11:43:40.44&-01$^{\circ}$43'54.98"&11:43:40.44&-01$^{\circ}$43'54.98"&0.19&14.23&02.99\\
(5817) Robertfrazer&j8vp10010&ACS/WFC&00:43:05.50&41$^{\circ}$24'58.37"&00:43:05.50&41$^{\circ}$24'58.37"&0.13&15.40&01.04\\
(941) Murray&ibom13010&WFC3/UVIS&17:54:38.94&-29$^{\circ}$48'49.09"&17:54:38.94&-29$^{\circ}$48'49.09"&0.10&15.64&01.92\\
(3236) Strand&jc6i02010&ACS/WFC&23:27:08.78&-02$^{\circ}$00'47.75"&23:27:08.78&-02$^{\circ}$00'47.75"&0.12&15.65&01.09\\
(5081) Sanguin&j8fs71qxq&ACS/WFC&12:16:33.12&13$^{\circ}$01'32.82"&12:16:33.12&13$^{\circ}$01'32.82"&0.96&16.63&01.70\\
(5081) Sanguin&j8fs71010&ACS/WFC&12:16:33.21&13$^{\circ}$01'30.90"&12:16:33.21&13$^{\circ}$01'30.90"&0.96&16.63&01.70\\
(5081) Sanguin&j8fs71020&ACS/WFC&12:16:33.50&13$^{\circ}$01'20.47"&12:16:33.50&13$^{\circ}$01'20.47"&0.96&16.63&01.70\\
(1275) Cimbria&jbz077010&ACS/WFC&11:33:29.84&03$^{\circ}$28'55.61"&11:33:29.84&03$^{\circ}$28'55.61"&0.18&16.65&03.17\\
(18886) 2000 AN164&j8pu1q010&ACS/WFC&10:01:28.86&02$^{\circ}$14'25.02"&10:01:28.86&02$^{\circ}$14'25.02"&1.03&16.74&01.45\\
(7247) Robertstirling&jcol34010&ACS/WFC&05:35:38.78&-05$^{\circ}$22'24.70"&05:35:38.78&-05$^{\circ}$22'24.70"&0.03&16.85&00.98\\
(16403) 1984 WJ1&j6jt04010&ACS/WFC&09:19:51.71&33$^{\circ}$45'04.07"&09:19:51.71&33$^{\circ}$45'04.07"&0.75&16.87&01.47\\
(16403) 1984 WJ1&j6jt04020&ACS/WFC&09:19:49.53&33$^{\circ}$45'05.29"&09:19:49.53&33$^{\circ}$45'05.29"&0.75&16.87&01.47\\
(25574) 1999 XZ205&icau74eyq&WFC3/UVIS&18:31:20.51&-32$^{\circ}$20'18.58"&18:31:20.51&-32$^{\circ}$20'18.58"&0.16&17.09&01.67\\
(25574) 1999 XZ205&icau74010&WFC3/UVIS&18:31:18.62&-32$^{\circ}$20'28.03"&18:31:18.62&-32$^{\circ}$20'28.03"&0.16&17.09&01.67\\
(25574) 1999 XZ205&icau74f1q&WFC3/UVIS&18:31:19.04&-32$^{\circ}$20'25.01"&18:31:19.04&-32$^{\circ}$20'25.01"&0.16&17.09&01.67\\
(1358) Gaika&j95420020&ACS/WFC&18:07:26.54&-24$^{\circ}$58'08.14"&18:07:26.54&-24$^{\circ}$58'08.14"&0.07&17.13&02.19\\
(1358) Gaika&j95420010&ACS/WFC&18:07:25.62&-24$^{\circ}$58'05.46"&18:07:25.62&-24$^{\circ}$58'05.46"&0.07&17.13&02.19\\
(1358) Gaika&j95420080&ACS/WFC&18:07:25.34&-24$^{\circ}$58'04.40"&18:07:25.34&-24$^{\circ}$58'04.40"&0.07&17.13&02.19\\
(11072) Hiraoka&u8l8f202m&WFPC2/WFC&13:06:34.20&03$^{\circ}$57'03.27"&13:06:34.20&03$^{\circ}$57'03.27"&1.03&17.14&01.52\\
(16974) Iphthime&ibyk08grq&WFC3/UVIS&08:07:18.86&03$^{\circ}$32'14.98"&08:07:18.86&03$^{\circ}$32'14.98"&0.08&17.34&04.64\\
(16974) Iphthime&ibyk08g1q&WFC3/UVIS&08:07:19.48&03$^{\circ}$32'05.87"&08:07:19.48&03$^{\circ}$32'05.87"&0.08&17.34&04.64\\
(16974) Iphthime&ibyk08gvq&WFC3/UVIS&08:07:18.29&03$^{\circ}$32'18.63"&08:07:18.29&03$^{\circ}$32'18.63"&0.08&17.34&04.64\\
(16974) Iphthime&ibyk08fzq&WFC3/UVIS&08:07:19.56&03$^{\circ}$32'05.08"&08:07:19.56&03$^{\circ}$32'05.08"&0.08&17.34&04.64\\
(16974) Iphthime&ibyk08gxq&WFC3/UVIS&08:07:18.24&03$^{\circ}$32'18.86"&08:07:18.24&03$^{\circ}$32'18.86"&0.08&17.34&04.64\\
(16974) Iphthime&ibyk08gmq&WFC3/UVIS&08:07:19.05&03$^{\circ}$32'08.06"&08:07:19.05&03$^{\circ}$32'08.06"&0.08&17.34&04.64\\
(16974) Iphthime&ibyk08gqq&WFC3/UVIS&08:07:18.88&03$^{\circ}$32'14.54"&08:07:18.88&03$^{\circ}$32'14.54"&0.08&17.34&04.64\\
(16974) Iphthime&ibyk08g0q&WFC3/UVIS&08:07:19.52&03$^{\circ}$32'05.50"&08:07:19.52&03$^{\circ}$32'05.50"&0.08&17.34&04.64\\
(16974) Iphthime&ibyk08guq&WFC3/UVIS&08:07:18.31&03$^{\circ}$32'18.54"&08:07:18.31&03$^{\circ}$32'18.54"&0.08&17.34&04.64\\
(16974) Iphthime&ibyk08gsq&WFC3/UVIS&08:07:18.83&03$^{\circ}$32'15.44"&08:07:18.83&03$^{\circ}$32'15.44"&0.08&17.34&04.64\\
(16974) Iphthime&ibyk08gnq&WFC3/UVIS&08:07:19.02&03$^{\circ}$32'08.17"&08:07:19.02&03$^{\circ}$32'08.17"&0.08&17.34&04.64\\
(16974) Iphthime&ibyk08goq&WFC3/UVIS&08:07:19.00&03$^{\circ}$32'08.28"&08:07:19.00&03$^{\circ}$32'08.28"&0.08&17.34&04.64\\
(16974) Iphthime&ibyk08fyq&WFC3/UVIS&08:07:19.59&03$^{\circ}$32'04.66"&08:07:19.59&03$^{\circ}$32'04.66"&0.08&17.34&04.64\\
(16974) Iphthime&ibyk08glq&WFC3/UVIS&08:07:19.08&03$^{\circ}$32'07.96"&08:07:19.08&03$^{\circ}$32'07.96"&0.08&17.34&04.64\\
(16974) Iphthime&ibyk08gwq&WFC3/UVIS&08:07:18.26&03$^{\circ}$32'18.75"&08:07:18.26&03$^{\circ}$32'18.75"&0.08&17.34&04.64\\
(16974) Iphthime&ibyk08gtq&WFC3/UVIS&08:07:18.79&03$^{\circ}$32'15.84"&08:07:18.79&03$^{\circ}$32'15.84"&0.08&17.34&04.64\\
(16594) Sorachi&jbts49010&ACS/WFC&17:59:01.43&-29$^{\circ}$11'53.54"&17:59:01.43&-29$^{\circ}$11'53.54"&0.08&17.37&01.99\\
(16594) Sorachi&jbts49020&ACS/WFC&17:59:01.60&-29$^{\circ}$11'55.08"&17:59:01.60&-29$^{\circ}$11'55.08"&0.08&17.37&01.99\\
(35843) 1999 JZ59&jcaj08010&ACS/WFC&17:49:12.57&-20$^{\circ}$24'42.23"&17:49:12.57&-20$^{\circ}$24'42.23"&0.66&17.42&01.07\\
(66575) 1999 RX152&jc6i01020&ACS/WFC&23:27:47.57&-02$^{\circ}$01'24.96"&23:27:47.57&-02$^{\circ}$01'24.96"&0.26&17.44&01.19\\
(136108) Haumea&j9fs20kdq&ACS/HRC&13:32:12.69&19$^{\circ}$23'36.80"&13:32:12.69&19$^{\circ}$23'36.80"&0.13&17.45&51.10\\
(136108) Haumea&j9fs20kgq&ACS/HRC&13:32:12.70&19$^{\circ}$23'37.85"&13:32:12.70&19$^{\circ}$23'37.85"&0.13&17.45&51.10\\
(11648) 1997 BT3&u9op5404m&WFPC2/WFC&13:18:34.00&-03$^{\circ}$13'45.07"&13:18:34.00&-03$^{\circ}$13'45.07"&0.86&17.48&01.86\\
(11648) 1997 BT3&u9op5403m&WFPC2/WFC&13:18:34.54&-03$^{\circ}$13'49.41"&13:18:34.54&-03$^{\circ}$13'49.41"&0.86&17.48&01.86\\
(3158) Anga&j6mf29010&ACS/WFC&14:06:49.23&-11$^{\circ}$22'01.81"&14:06:49.23&-11$^{\circ}$22'01.81"&0.05&17.64&02.43\\
(23318)Salvadorsanchez&jck905010&ACS/WFC&18:18:27.41&-13$^{\circ}$42'27.04"&18:18:27.41&-13$^{\circ}$42'27.04"&0.06&17.67&02.45\\
(23318)Salvadorsanchez&jck905020&ACS/WFC&18:18:27.42&-13$^{\circ}$42'34.95"&18:18:27.42&-13$^{\circ}$42'34.95"&0.06&17.67&02.45\\
(23318)Salvadorsanchez&jck907020&ACS/WFC&18:18:41.01&-13$^{\circ}$46'39.30"&18:18:41.01&-13$^{\circ}$46'39.30"&0.06&17.68&02.46\\
(11616) 1996 BQ2&j9s955010&ACS/WFC&02:17:55.70&-01$^{\circ}$13'12.95"&02:17:55.70&-01$^{\circ}$13'12.95"&0.91&17.68&02.87\\
(23318)Salvadorsanchez&jck908010&ACS/WFC&18:18:44.50&-13$^{\circ}$47'32.26"&18:18:44.50&-13$^{\circ}$47'32.26"&0.06&17.68&02.46\\
(23318)Salvadorsanchez&jck907010&ACS/WFC&18:18:40.98&-13$^{\circ}$46'31.39"&18:18:40.98&-13$^{\circ}$46'31.39"&0.06&17.68&02.46\\
(9664) Brueghel&n43h05b7q&NICMOS/NIC2&20:37:23.83&-19$^{\circ}$12'16.90"&20:37:23.83&-19$^{\circ}$12'16.90"&0.29&17.71&02.24\\
(10226) Seishika&jchx51010&ACS/WFC&06:08:45.68&20$^{\circ}$34'11.86"&06:08:45.68&20$^{\circ}$34'11.86"&0.06&17.71&02.15\\
(9664) Brueghel&n43h05b8q&NICMOS/NIC2&20:37:23.70&-19$^{\circ}$12'16.92"&20:37:23.70&-19$^{\circ}$12'16.92"&0.29&17.71&02.24\\
(9664) Brueghel&n43h05b3q&NICMOS/NIC2&20:37:24.12&-19$^{\circ}$12'16.68"&20:37:24.12&-19$^{\circ}$12'16.68"&0.29&17.72&02.24\\
(2042) Sitarski&id3i01020&WFC3/UVIS&02:32:31.72&18$^{\circ}$36'03.24"&02:32:31.72&18$^{\circ}$36'03.24"&0.12&17.72&02.64\\
(26499) 2000 CX1&jc2910010&ACS/WFC&10:44:31.50&12$^{\circ}$07'33.73"&10:44:31.50&12$^{\circ}$07'33.73"&0.34&17.72&01.93\\
(9664) Brueghel&n43h05b5q&NICMOS/NIC2&20:37:23.99&-19$^{\circ}$12'16.81"&20:37:23.99&-19$^{\circ}$12'16.81"&0.29&17.72&02.24\\
(28227) 1999 AN2&j8zb08010&ACS/WFC&23:49:36.22&-09$^{\circ}$37'16.55"&23:49:36.22&-09$^{\circ}$37'16.55"&0.64&17.72&01.47\\
(2683) Brian&ibi901020&WFC3/UVIS&09:47:48.79&13$^{\circ}$17'23.39"&09:47:48.79&13$^{\circ}$17'23.39"&0.63&17.75&03.25\\
(6659) Pietsch&j8mbi8boq&ACS/WFC&06:09:25.62&24$^{\circ}$28'40.42"&06:09:25.62&24$^{\circ}$28'40.42"&0.05&17.83&02.31\\
(6659) Pietsch&j8mbi8bmq&ACS/WFC&06:09:24.84&24$^{\circ}$28'40.77"&06:09:24.84&24$^{\circ}$28'40.77"&0.05&17.83&02.31\\
(6659) Pietsch&j8mbi8bjq&ACS/WFC&06:09:23.26&24$^{\circ}$28'38.31"&06:09:23.26&24$^{\circ}$28'38.31"&0.05&17.83&02.31\\
(6659) Pietsch&j8mbi8bkq&ACS/WFC&06:09:24.10&24$^{\circ}$28'39.69"&06:09:24.10&24$^{\circ}$28'39.69"&0.05&17.83&02.31\\
(20731) Mothediniz&j9bl02050&ACS/WFC&06:59:55.00&14$^{\circ}$15'16.66"&06:59:55.00&14$^{\circ}$15'16.66"&0.63&17.88&02.06\\
(20731) Mothediniz&j9bl02040&ACS/WFC&06:59:56.15&14$^{\circ}$15'11.29"&06:59:56.15&14$^{\circ}$15'11.29"&0.63&17.89&02.06\\
(20731) Mothediniz&j9bl02030&ACS/WFC&06:59:57.20&14$^{\circ}$15'10.58"&06:59:57.20&14$^{\circ}$15'10.58"&0.63&17.89&02.06\\
(3149) Okudzhava&ib6w38010&WFC3/UVIS&12:39:09.28&-00$^{\circ}$33'20.73"&12:39:09.28&-00$^{\circ}$33'20.73"&0.59&17.91&01.73\\
(3149) Okudzhava&ib6w38020&WFC3/UVIS&12:39:08.47&-00$^{\circ}$33'04.76"&12:39:08.47&-00$^{\circ}$33'04.76"&0.59&17.92&01.73\\
(109640) 2001 RJ&jbf407010&ACS/WFC&00:45:15.58&41$^{\circ}$58'28.14"&00:45:15.58&41$^{\circ}$58'28.14"&0.72&17.93&01.78\\
(68812) 2002 GB56&icii81h5q&WFC3/UVIS&18:43:36.56&-21$^{\circ}$00'58.56"&18:43:36.56&-21$^{\circ}$00'58.56"&0.13&17.96&01.53\\
(68812) 2002 GB56&icii82hjq&WFC3/UVIS&18:43:31.41&-21$^{\circ}$01'02.58"&18:43:31.41&-21$^{\circ}$01'02.58"&0.13&17.96&01.53\\
(68812) 2002 GB56&icii82hiq&WFC3/UVIS&18:43:31.75&-21$^{\circ}$00'59.53"&18:43:31.75&-21$^{\circ}$00'59.53"&0.13&17.96&01.53\\
(68812) 2002 GB56&icii81h7q&WFC3/UVIS&18:43:36.04&-21$^{\circ}$00'58.34"&18:43:36.04&-21$^{\circ}$00'58.34"&0.13&17.96&01.53\\
(68812) 2002 GB56&icii81h3q&WFC3/UVIS&18:43:37.09&-21$^{\circ}$00'57.25"&18:43:37.09&-21$^{\circ}$00'57.25"&0.13&17.96&01.53\\
(2883) Barabashov&icpg20obq&WFC3/UVIS&17:37:32.03&-24$^{\circ}$22'27.29"&17:37:32.03&-24$^{\circ}$22'27.29"&0.21&17.97&02.58\\
(2883) Barabashov&icpg20ogq&WFC3/UVIS&17:37:33.15&-24$^{\circ}$22'29.14"&17:37:33.15&-24$^{\circ}$22'29.14"&0.21&17.97&02.58\\
(2883) Barabashov&icpg20olq&WFC3/UVIS&17:37:33.81&-24$^{\circ}$22'29.43"&17:37:33.81&-24$^{\circ}$22'29.43"&0.21&17.97&02.58\\
(2883) Barabashov&icpg20omq&WFC3/UVIS&17:37:33.95&-24$^{\circ}$22'29.43"&17:37:33.95&-24$^{\circ}$22'29.43"&0.21&17.97&02.58\\
(2883) Barabashov&icpg20ohq&WFC3/UVIS&17:37:33.26&-24$^{\circ}$22'29.22"&17:37:33.26&-24$^{\circ}$22'29.22"&0.21&17.97&02.58\\
(2883) Barabashov&icpg20o9q&WFC3/UVIS&17:37:31.78&-24$^{\circ}$22'26.75"&17:37:31.78&-24$^{\circ}$22'26.75"&0.21&17.97&02.58\\
(2883) Barabashov&icpg20nwq&WFC3/UVIS&17:37:30.23&-24$^{\circ}$22'24.19"&17:37:30.23&-24$^{\circ}$22'24.19"&0.21&17.97&02.58\\
(2883) Barabashov&icpg20oiq&WFC3/UVIS&17:37:33.38&-24$^{\circ}$22'29.30"&17:37:33.38&-24$^{\circ}$22'29.30"&0.21&17.97&02.58\\
(2883) Barabashov&icpg20nyq&WFC3/UVIS&17:37:30.48&-24$^{\circ}$22'24.43"&17:37:30.48&-24$^{\circ}$22'24.43"&0.21&17.97&02.58\\
(2883) Barabashov&icpg20nzq&WFC3/UVIS&17:37:30.61&-24$^{\circ}$22'24.59"&17:37:30.61&-24$^{\circ}$22'24.59"&0.21&17.97&02.58\\
(2883) Barabashov&icpg20o1q&WFC3/UVIS&17:37:30.85&-24$^{\circ}$22'24.91"&17:37:30.85&-24$^{\circ}$22'24.91"&0.21&17.97&02.58\\
(2883) Barabashov&icpg20o7q&WFC3/UVIS&17:37:31.58&-24$^{\circ}$22'26.30"&17:37:31.58&-24$^{\circ}$22'26.30"&0.21&17.97&02.58\\
(2883) Barabashov&icpg20o8q&WFC3/UVIS&17:37:31.67&-24$^{\circ}$22'26.50"&17:37:31.67&-24$^{\circ}$22'26.50"&0.21&17.97&02.58\\
(2883) Barabashov&icpg20o4q&WFC3/UVIS&17:37:31.26&-24$^{\circ}$22'25.63"&17:37:31.26&-24$^{\circ}$22'25.63"&0.21&17.97&02.58\\
(2883) Barabashov&icpg20o2q&WFC3/UVIS&17:37:30.95&-24$^{\circ}$22'25.08"&17:37:30.95&-24$^{\circ}$22'25.08"&0.21&17.97&02.58\\
(2883) Barabashov&icpg20nxq&WFC3/UVIS&17:37:30.37&-24$^{\circ}$22'24.32"&17:37:30.37&-24$^{\circ}$22'24.32"&0.21&17.97&02.58\\
(2883) Barabashov&icpg20ocq&WFC3/UVIS&17:37:32.15&-24$^{\circ}$22'27.56"&17:37:32.15&-24$^{\circ}$22'27.56"&0.21&17.97&02.58\\
(2883) Barabashov&icpg20o5q&WFC3/UVIS&17:37:31.37&-24$^{\circ}$22'25.86"&17:37:31.37&-24$^{\circ}$22'25.86"&0.21&17.97&02.58\\
(2883) Barabashov&icpg20oaq&WFC3/UVIS&17:37:31.88&-24$^{\circ}$22'26.96"&17:37:31.88&-24$^{\circ}$22'26.96"&0.21&17.97&02.58\\
(2883) Barabashov&icpg20oeq&WFC3/UVIS&17:37:32.94&-24$^{\circ}$22'28.91"&17:37:32.94&-24$^{\circ}$22'28.91"&0.21&17.97&02.58\\
(2883) Barabashov&icpg20ofq&WFC3/UVIS&17:37:33.04&-24$^{\circ}$22'29.02"&17:37:33.04&-24$^{\circ}$22'29.02"&0.21&17.97&02.58\\
(2883) Barabashov&icpg20o0q&WFC3/UVIS&17:37:30.72&-24$^{\circ}$22'24.73"&17:37:30.72&-24$^{\circ}$22'24.73"&0.21&17.97&02.58\\
(2883) Barabashov&icpg20o6q&WFC3/UVIS&17:37:31.47&-24$^{\circ}$22'26.05"&17:37:31.47&-24$^{\circ}$22'26.05"&0.21&17.97&02.58\\
(2883) Barabashov&icpg20okq&WFC3/UVIS&17:37:33.66&-24$^{\circ}$22'29.41"&17:37:33.66&-24$^{\circ}$22'29.41"&0.21&17.97&02.58\\
(2883) Barabashov&icpg20o3q&WFC3/UVIS&17:37:31.12&-24$^{\circ}$22'25.37"&17:37:31.12&-24$^{\circ}$22'25.37"&0.21&17.97&02.58\\
\end{longtable}
\tablefoot{Removed from this list serendipitous detections of (134340) Pluto where the observation target was its moon Charon I.\\
\tablefoottext{a}{Predicted Right Ascension (RA, J2000) and Declination (Dec, J2000) at the start of the observation}
\tablefoottext{b}{Predicted Right Ascension (RA, J2000) and Declination (Dec, J2000) at the end of the exposure time}
\tablefoottext{c}{Propagated position uncertainty}
\tablefoottext{d}{Distance from the satellite at the time of the observation}}
}
%----------------------------------------------------------------------------------------------------%
%----------------------------------------------------------------------------------------------------%
\bigskip % just to make a space here
%----------------------------------------------------------------------------------------------------%
%----------------------------------------------------------------------------------------------------%
\longtab[2]{    
\begin{longtable}{p{2.4cm}ccccccccc}
\caption{{\it XMM-Newton Serendipitous Detections of Asteroids}. A sample of the first 100 serendipitous detections of asteroids from the OM instrument, selected with $\delta$Pos < 5 arcsec, cross-match type 2, and with a theoretical apparent magnitude above the limiting magnitude per observation and filter.}
\label{table:xmmXmatch_list}\\
\hline
\hline
Asteroid Id & Observation Id & Filter &  RA$_1$ & Dec$_1$\tablefootmark{a} & RA$_2$ & Dec$_2$\tablefootmark{b} & $\delta$Pos\tablefootmark{c}&  m$_{v}\_zeropoint$\tablefootmark{d} & d\tablefootmark{e}\\
 & & & \multicolumn{2}{c}{\it (J2000.0)}&\multicolumn{2}{c}{\it (J2000.0)} & \it{(arcsec)} & & \it{(AU)}\\
\hline
\\
\endfirsthead
\caption{Continued.} \\
\hline
Asteroid Id & Observation Id & Filter &  RA$_1$ & Dec$_1$\tablefootmark{a} & RA$_2$ & Dec$_2$\tablefootmark{b} & $\delta$Pos\tablefootmark{c}&  m$_{v}\_zeropoint$\tablefootmark{d} & d\tablefootmark{e}\\
 & & & \multicolumn{2}{c}{\it (J2000.0)}&\multicolumn{2}{c}{\it (J2000.0)} & \it{(arcsec)} & & \it{(AU)}\\
\hline
\\
\endhead
\hline
\endfoot
\hline
\\
\endlastfoot
(234) Barbara&0781040101&U&03:20:02.81&00$^{\circ}$27'48.18"&03:20:02.81&00$^{\circ}$27'48.18"&0.12&13.10&01.34\\
(234) Barbara&0781040101&L&03:20:02.81&00$^{\circ}$27'48.18"&03:20:02.81&00$^{\circ}$27'48.18"&0.12&13.84&01.34\\
(386) Siegena&0694510101&B&14:08:36.77&-03$^{\circ}$05'11.66"&14:08:36.77&-03$^{\circ}$05'11.66"&0.27&14.29&03.55\\
(386) Siegena&0694510101&U&14:08:36.77&-03$^{\circ}$05'11.66"&14:08:36.77&-03$^{\circ}$05'11.66"&0.27&14.43&03.55\\
(384) Burdigala&0650590401&B&12:23:07.96&02$^{\circ}$50'43.42"&12:23:07.96&02$^{\circ}$50'43.42"&0.12&15.41&02.62\\
(624) Hektor&0165972001&B&18:57:10.71&-37$^{\circ}$46'20.30"&18:57:10.71&-37$^{\circ}$46'20.30"&0.28&15.70&05.03\\
(386) Siegena&0694510101&L&14:08:36.77&-03$^{\circ}$05'11.66"&14:08:36.77&-03$^{\circ}$05'11.66"&0.27&15.17&03.55\\
(892) Seeligeria&0303670101&B&14:20:16.56&06$^{\circ}$39'28.61"&14:20:16.56&06$^{\circ}$39'28.61"&0.50&16.23&02.99\\
(1107) Lictoria&0012440101&B&00:55:59.04&-01$^{\circ}$24'55.14"&00:55:59.04&-01$^{\circ}$24'55.14"&0.10&15.89&03.08\\
(5619) Shair&0694580101&U&22:19:00.97&12$^{\circ}$04'01.00"&22:19:00.97&12$^{\circ}$04'01.00"&0.14&17.64&01.86\\
(624) Hektor&0165972001&U&18:57:10.71&-37$^{\circ}$46'20.30"&18:57:10.71&-37$^{\circ}$46'20.30"&0.28&15.84&05.03\\
(90075) 2002 VU94&0800400601&U&08:50:46.33&-00$^{\circ}$03'10.91"&08:50:46.33&-00$^{\circ}$03'10.91"&0.01&16.19&00.37\\
(624) Hektor&0165972001&V&18:57:10.71&-37$^{\circ}$46'20.30"&18:57:10.71&-37$^{\circ}$46'20.30"&0.28&15.15&05.03\\
(2679) Kittisvaara&0205570101&B&03:39:17.64&10$^{\circ}$00'25.52"&03:39:17.64&10$^{\circ}$00'25.52"&0.76&17.21&02.06\\
(677) Aaltje&0302352401&L&09:59:01.06&02$^{\circ}$02'15.46"&09:59:01.06&02$^{\circ}$02'15.46"&0.25&16.28&02.44\\
(234) Barbara&0781040101&M&03:20:02.81&00$^{\circ}$27'48.18"&03:20:02.81&00$^{\circ}$27'48.18"&0.12&15.97&01.34\\
(1137) Raissa&0763100101&U&18:09:36.25&-26$^{\circ}$06'28.69"&18:09:36.25&-26$^{\circ}$06'28.69"&0.37&16.10&02.30\\
(892) Seeligeria&0303670101&U&14:20:16.56&06$^{\circ}$39'28.61"&14:20:16.56&06$^{\circ}$39'28.61"&0.50&16.37&02.99\\
(892) Seeligeria&0303670101&V&14:20:16.56&06$^{\circ}$39'28.61"&14:20:16.56&06$^{\circ}$39'28.61"&0.50&15.68&02.99\\
(865) Zubaida&0405210101&V&05:29:46.52&12$^{\circ}$11'15.76"&05:29:46.52&12$^{\circ}$11'15.76"&0.54&16.31&02.18\\
(1137) Raissa&0763100101&L&18:09:36.25&-26$^{\circ}$06'28.69"&18:09:36.25&-26$^{\circ}$06'28.69"&0.37&16.84&02.30\\
(1908) Pobeda&0305540501&L&16:26:48.24&-24$^{\circ}$45'36.58"&16:26:48.24&-24$^{\circ}$45'36.58"&0.33&17.86&02.45\\
(5293)Bentengahama&0109661401&B&01:44:23.71&-04$^{\circ}$30'27.62"&01:44:23.71&-04$^{\circ}$30'27.62"&0.83&17.73&02.57\\
(3784) Chopin&0723802101&B&00:40:35.10&-09$^{\circ}$28'21.09"&00:40:35.10&-09$^{\circ}$28'21.09"&0.12&17.48&02.78\\
(715) Transvaalia&0821871601&L&22:48:04.51&-22$^{\circ}$12'54.73"&22:48:04.51&-22$^{\circ}$12'54.73"&0.03&16.70&02.63\\
(1505) Koranna&0301651201&U&23:05:03.36&12$^{\circ}$15'04.16"&23:05:03.36&12$^{\circ}$15'04.16"&0.14&17.14&02.29\\
(2464) Nordenskiold&0672720101&B&08:09:19.91&20$^{\circ}$50'06.73"&08:09:19.91&20$^{\circ}$50'06.73"&0.24&17.69&02.47\\
(5010) Amenemhet&0556212301&B&14:54:41.99&01$^{\circ}$43'54.90"&14:54:41.99&01$^{\circ}$43'54.90"&0.19&17.32&01.80\\
(1333) Cevenola&0723801601&B&23:25:20.55&-12$^{\circ}$00'16.78"&23:25:20.55&-12$^{\circ}$00'16.78"&0.04&17.62&02.84\\
(13859) Fredtreasure&0303561001&U&12:50:20.35&-23$^{\circ}$32'20.02"&12:50:20.35&-23$^{\circ}$32'20.02"&0.80&17.17&01.73\\
(14425) Fujimimachi&0555220101&U&21:59:02.96&-20$^{\circ}$04'15.88"&21:59:02.96&-20$^{\circ}$04'15.88"&0.26&17.53&01.19\\
(1860) Barbarossa&0747400101&L&01:11:49.36&-00$^{\circ}$19'12.92"&01:11:49.36&-00$^{\circ}$19'12.92"&0.04&17.26&01.78\\
(3784) Chopin&0723802101&V&00:40:35.10&-09$^{\circ}$28'21.09"&00:40:35.10&-09$^{\circ}$28'21.09"&0.12&16.93&02.78\\
(6406) Mikejura&0205570101&B&03:39:17.46&10$^{\circ}$08'55.70"&03:39:17.46&10$^{\circ}$08'55.70"&0.64&18.27&01.79\\
(5293)Bentengahama&0109661401&U&01:44:23.71&-04$^{\circ}$30'27.62"&01:44:23.71&-04$^{\circ}$30'27.62"&0.83&17.87&02.57\\
(3784) Chopin&0723802101&U&00:40:35.10&-09$^{\circ}$28'21.09"&00:40:35.10&-09$^{\circ}$28'21.09"&0.12&17.62&02.78\\
(90075) 2002 VU94&0800400601&L&08:50:46.33&-00$^{\circ}$03'10.91"&08:50:46.33&-00$^{\circ}$03'10.91"&0.01&16.93&00.37\\
(892) Seeligeria&0303670101&L&14:20:16.56&06$^{\circ}$39'28.61"&14:20:16.56&06$^{\circ}$39'28.61"&0.50&17.11&02.99\\
(7468) Anfimov&0790800101&U&23:39:29.59&-05$^{\circ}$37'26.94"&23:39:29.59&-05$^{\circ}$37'26.94"&0.08&18.33&02.64\\
(3021) Lucubratio&0110980101&B&11:20:14.78&13$^{\circ}$40'07.12"&11:20:14.78&13$^{\circ}$40'07.12"&0.71&17.74&02.54\\
(2118) Flagstaff&0804250301&B&17:46:37.34&-29$^{\circ}$49'55.56"&17:46:37.34&-29$^{\circ}$49'55.56"&0.07&18.11&03.02\\
(31354) 1998 TR3&0744390201&U&00:19:01.85&-20$^{\circ}$29'24.59"&00:19:01.85&-20$^{\circ}$29'24.59"&0.08&18.62&01.77\\
(1345) Potomac&0674480401&U&14:04:14.95&-01$^{\circ}$47'41.07"&14:04:14.95&-01$^{\circ}$47'41.07"&0.11&16.97&03.58\\
(1860) Barbarossa&0747390101&L&01:03:25.83&-00$^{\circ}$21'23.06"&01:03:25.83&-00$^{\circ}$21'23.06"&0.04&17.36&01.86\\
(6306) Nishimura&0112370801&B&02:19:06.66&-05$^{\circ}$12'54.76"&02:19:06.66&-05$^{\circ}$12'54.76"&1.08&17.68&01.88\\
(2464) Nordenskiold&0672720101&U&08:09:19.91&20$^{\circ}$50'06.73"&08:09:19.91&20$^{\circ}$50'06.73"&0.24&17.83&02.47\\
(1331) Solvejg&0611181501&L&05:34:41.17&21$^{\circ}$59'21.75"&05:34:41.17&21$^{\circ}$59'21.75"&0.16&17.72&03.15\\
(1747) Wright&0741891201&L&18:12:28.62&00$^{\circ}$38'42.53"&18:12:28.62&00$^{\circ}$38'42.53"&0.09&17.44&01.03\\
(2641) Lipschutz&0744440301&U&12:46:20.93&02$^{\circ}$31'26.96"&12:46:20.93&02$^{\circ}$31'26.96"&0.23&17.76&01.98\\
(2641) Lipschutz&0744440301&V&12:46:20.93&02$^{\circ}$31'26.96"&12:46:20.93&02$^{\circ}$31'26.96"&0.23&17.07&01.98\\
(1333) Cevenola&0723801601&V&23:25:20.55&-12$^{\circ}$00'16.78"&23:25:20.55&-12$^{\circ}$00'16.78"&0.04&17.07&02.84\\
(14376) 1989 ST10&0721010501&B&17:28:13.36&-14$^{\circ}$11'07.14"&17:28:13.36&-14$^{\circ}$11'07.14"&0.02&18.11&01.88\\
(6446) Lomberg&0727571501&V&03:53:56.12&-00$^{\circ}$10'42.36"&03:53:56.12&-00$^{\circ}$10'42.36"&0.08&18.33&01.34\\
(93719) 2000 VR36&0821250601&U&02:23:20.34&-03$^{\circ}$30'44.93"&02:23:20.34&-03$^{\circ}$30'44.93"&0.02&18.78&01.99\\
(2381) Landi&0700182001&U&21:29:05.29&-07$^{\circ}$47'37.47"&21:29:05.29&-07$^{\circ}$47'37.47"&0.28&17.96&02.82\\
(4358) Lynn&0744490601&B&22:36:33.06&-12$^{\circ}$46'06.41"&22:36:33.06&-12$^{\circ}$46'06.41"&0.09&17.86&02.52\\
(19878) 1030 T-1&0110980601&B&13:19:34.22&-14$^{\circ}$51'21.10"&13:19:34.22&-14$^{\circ}$51'21.10"&0.42&18.61&01.47\\
(10823) Sakaguchi&0748190101&B&01:56:44.55&05$^{\circ}$43'22.30"&01:56:44.55&05$^{\circ}$43'22.30"&0.09&18.57&01.47\\
(3493) Stepanov&0081340801&B&12:13:24.64&02$^{\circ}$42'19.38"&12:13:24.64&02$^{\circ}$42'19.38"&1.38&17.67&01.70\\
(2223) Sarpedon&0800400501&U&09:10:03.44&-00$^{\circ}$46'01.63"&09:10:03.44&-00$^{\circ}$46'01.63"&0.08&17.86&04.87\\
(1107) Lictoria&0012440101&L&00:55:59.04&-01$^{\circ}$24'55.14"&00:55:59.04&-01$^{\circ}$24'55.14"&0.10&16.77&03.08\\
(4225) Hobart&0652350801&U&16:30:12.61&-24$^{\circ}$37'26.06"&16:30:12.61&-24$^{\circ}$37'26.06"&0.09&17.46&01.52\\
(4024) Ronan&0782520501&L&10:22:20.58&19$^{\circ}$51'41.99"&10:22:20.58&19$^{\circ}$51'41.99"&0.07&18.22&01.51\\
(2677) Joan&0601391001&B&20:44:26.22&-10$^{\circ}$47'59.12"&20:44:26.22&-10$^{\circ}$47'59.12"&0.33&17.95&03.15\\
(234) Barbara&0781040101&S&03:20:02.81&00$^{\circ}$27'48.18"&03:20:02.81&00$^{\circ}$27'48.18"&0.12&16.44&01.34\\
(523) Ada&0803030301&L&17:09:41.09&-24$^{\circ}$13'59.57"&17:09:41.09&-24$^{\circ}$13'59.57"&0.09&17.66&03.55\\
(5822) Masakichi&0203540401&U&04:33:03.94&24$^{\circ}$12'38.40"&04:33:03.94&24$^{\circ}$12'38.40"&0.35&18.25&01.61\\
(1295) Deflotte&0744500301&L&04:21:29.72&19$^{\circ}$28'55.69"&04:21:29.72&19$^{\circ}$28'55.69"&0.18&18.03&03.05\\
(6723) Chrisclark&0670120401&B&22:54:21.12&-17$^{\circ}$35'04.72"&22:54:21.12&-17$^{\circ}$35'04.72"&0.07&18.73&02.85\\
(1331) Solvejg&0611181401&L&05:34:37.25&21$^{\circ}$59'10.20"&05:34:37.25&21$^{\circ}$59'10.20"&0.16&17.72&03.14\\
(13859) Fredtreasure&0303561001&L&12:50:20.35&-23$^{\circ}$32'20.02"&12:50:20.35&-23$^{\circ}$32'20.02"&0.80&17.91&01.73\\
(12827) 1997 AS7&0653040101&B&07:51:10.35&17$^{\circ}$38'13.35"&07:51:10.35&17$^{\circ}$38'13.35"&0.60&19.36&02.23\\
(6723) Chrisclark&0670120301&B&22:54:05.75&-17$^{\circ}$49'29.43"&22:54:05.75&-17$^{\circ}$49'29.43"&0.07&18.71&02.82\\
(17403) Masciarelli&0402430301&U&17:45:19.83&-28$^{\circ}$56'09.26"&17:45:19.83&-28$^{\circ}$56'09.26"&1.00&18.96&02.31\\
(8573) Ivanka&0692330401&U&07:28:45.43&33$^{\circ}$53'40.40"&07:28:45.43&33$^{\circ}$53'40.40"&0.30&18.40&02.39\\
(3021) Lucubratio&0110980701&U&11:20:06.55&13$^{\circ}$40'54.86"&11:20:06.55&13$^{\circ}$40'54.86"&0.71&17.88&02.54\\
(2918) Salazar&0803950801&B&09:47:38.18&14$^{\circ}$22'19.69"&09:47:38.18&14$^{\circ}$22'19.69"&0.21&18.66&03.11\\
(1623) Vivian&0693990301&L&11:18:59.66&06$^{\circ}$55'35.08"&11:18:59.66&06$^{\circ}$55'35.08"&0.07&18.88&03.32\\
(3051) Nantong&0552860101&U&10:54:33.71&-05$^{\circ}$40'15.29"&10:54:33.71&-05$^{\circ}$40'15.29"&0.62&18.39&02.54\\
(5122) Mucha&0300240101&B&09:18:36.19&16$^{\circ}$19'51.10"&09:18:36.19&16$^{\circ}$19'51.10"&0.32&18.22&02.75\\
(6306) Nishimura&0112370801&V&02:19:06.66&-05$^{\circ}$12'54.76"&02:19:06.66&-05$^{\circ}$12'54.76"&1.08&17.13&01.88\\
(5593) Jonsujatha&0555970401&B&14:56:12.61&-11$^{\circ}$31'33.75"&14:56:12.61&-11$^{\circ}$31'33.75"&0.67&18.76&02.41\\
(2118) Flagstaff&0804250301&U&17:46:37.34&-29$^{\circ}$49'55.56"&17:46:37.34&-29$^{\circ}$49'55.56"&0.07&18.25&03.02\\
(61312) 2000 OS50&0506050201&B&17:33:21.25&-24$^{\circ}$19'51.76"&17:33:21.25&-24$^{\circ}$19'51.76"&0.22&18.67&01.46\\
(6679) Gurzhij&0744490401&B&22:35:31.21&-12$^{\circ}$46'46.60"&22:35:31.21&-12$^{\circ}$46'46.60"&0.04&18.86&02.04\\
(14376) 1989 ST10&0721010501&V&17:28:13.36&-14$^{\circ}$11'07.14"&17:28:13.36&-14$^{\circ}$11'07.14"&0.02&17.56&01.88\\
(6306) Nishimura&0112370801&U&02:19:06.66&-05$^{\circ}$12'54.76"&02:19:06.66&-05$^{\circ}$12'54.76"&1.08&17.82&01.88\\
(3876) Quaide&0692510201&V&09:28:43.61&18$^{\circ}$49'52.22"&09:28:43.61&18$^{\circ}$49'52.22"&0.25&17.79&03.43\\
(1746) Brouwer&0801681301&L&17:49:31.46&-28$^{\circ}$31'49.71"&17:49:31.46&-28$^{\circ}$31'49.71"&0.08&18.07&03.89\\
(13025) Zurich&0674370201&U&09:07:11.86&14$^{\circ}$52'13.96"&09:07:11.86&14$^{\circ}$52'13.96"&0.46&18.37&01.88\\
(14376) 1989 ST10&0721010501&U&17:28:13.36&-14$^{\circ}$11'07.14"&17:28:13.36&-14$^{\circ}$11'07.14"&0.02&18.25&01.88\\
(8189) Naruke&0503490201&U&22:42:18.93&-09$^{\circ}$41'49.86"&22:42:18.93&-09$^{\circ}$41'49.86"&0.36&19.55&03.47\\
(5509) Rennsteig&0204580301&V&18:33:51.46&-21$^{\circ}$05'44.58"&18:33:51.46&-21$^{\circ}$05'44.58"&0.07&18.51&01.73\\
(190677) 2001 BQ61&0205670101&U&04:53:00.80&-02$^{\circ}$51'31.18"&04:53:00.80&-02$^{\circ}$51'31.18"&0.51&19.27&01.44\\
(2190) Coubertin&0153450101&B&12:47:09.60&-05$^{\circ}$42'52.21"&12:47:09.60&-05$^{\circ}$42'52.21"&0.39&18.35&02.07\\
(8956) 1998 FN119&0823360101&B&21:50:01.62&-05$^{\circ}$57'20.37"&21:50:01.62&-05$^{\circ}$57'20.37"&0.04&20.53&02.64\\
(89229) 2001 UY127&0810600201&U&03:20:59.29&-01$^{\circ}$09'29.92"&03:20:59.29&-01$^{\circ}$09'29.92"&0.07&19.06&01.98\\
(4358) Lynn&0744490601&U&22:36:33.06&-12$^{\circ}$46'06.41"&22:36:33.06&-12$^{\circ}$46'06.41"&0.09&18.00&02.52\\
(11272) 1988 RK&0744420101&B&18:26:00.12&-12$^{\circ}$59'14.79"&18:26:00.12&-12$^{\circ}$59'14.79"&0.06&19.06&02.15\\
(46573) 1992 AJ1&0202680101&U&17:33:36.45&-26$^{\circ}$05'27.17"&17:33:36.45&-26$^{\circ}$05'27.17"&0.09&18.97&02.29\\

\end{longtable}
\tablefoot{The results above are ordered by the greatest difference between the $m_{v}\_zeropoint$\tablefootmark{d} and $mlim\_obs$\tablefootmark{f}, both columns provided in the final catalogue.\\
\tablefoottext{a}{Predicted Right Ascension (RA, J2000) and Declination (Dec, J2000) at the start of the observation}
\tablefoottext{b}{Predicted Right Ascension (RA, J2000) and Declination (Dec, J2000) at the end of the exposure time}
\tablefoottext{c}{Propagated position uncertainty}
\tablefoottext{d}{zero-point magnitude corrected for the corresponding instrument filter}
\tablefoottext{e}{Distance from the satellite at the time of the observation}
\tablefoottext{f}{Limiting magnitude per observation\_id and filter retreived from the XMM-Newton pipeline products as described in Sec.\ref{sec:results}}
}
}

\end{appendix}

\end{document}